\documentclass[aip,amsmath,amssymb,reprint]{revtex4-2}

\usepackage{graphicx}
\usepackage{dcolumn}
\usepackage{bm}
\usepackage{booktabs}
\usepackage{setspace}
\usepackage{amsmath}
\usepackage{mathtools}
\usepackage{amssymb}
\usepackage{chngcntr}
\usepackage{tikz}
\usepackage[titletoc,toc,title]{appendix}

\usepackage{lineno}

\graphicspath{{./figures/}}

\usepackage{graphicx}
\usepackage{dcolumn}
\usepackage{bm}

\usepackage[utf8]{inputenc}
\usepackage[T1]{fontenc}
\usepackage{mathptmx}
\usepackage{etoolbox}

\usepackage{booktabs}
\usepackage{setspace}
\usepackage{amsmath}
\usepackage{mathtools}
\usepackage{amssymb}
\usepackage{chngcntr}
\usepackage{tikz}
\usepackage[titletoc,toc,title]{appendix}
\usepackage{lineno}

\makeatletter
\def\@email#1#2{%
 \endgroup
 \patchcmd{\titleblock@produce}
  {\frontmatter@RRAPformat}
  {\frontmatter@RRAPformat{\produce@RRAP{*#1\href{mailto:#2}{#2}}}\frontmatter@RRAPformat}
  {}{}
}%
\makeatother
\begin{document}


\preprint{AIP/123-QED}

\title[Galilean and quasi-Galilean Frames for Tidal Bores]{A Multi-frame Analysis of Flow Reversal in Tidal Bore}

\author{Assia Belemkadem}
\author{Johan Fourdrinoy}
\author{Anthony Beaudoin}
\author{Germain Rousseaux}
\affiliation{Institut Pprime (UPR 3346), CNRS Universit\'{e} de Poitiers ISAE ENSMA, 11 Boulevard Marie et Pierre Curie, T\'{e}l\'{e}port 2, BP 30179, 86962 Futuroscope Chasseneuil Cedex, France}
\homepage{https://pprime.fr}
\email{anthony.beaudoin@univ-poitiers.fr}

\date{\today}





\begin{abstract}
A tidal bore is a nonlinear step like surge propagating on the top of a river counter-current. The scientific literature reports that the passage of the tidal front may induce local flow reversals and, in some cases, a global reversal of the river current. Here, combining Particle Image Velocimetry (PIV) measurements and 2D (Two Dimensions) numerical simulations based on a Weakly Compressible Smoothed Particle Hydrodynamics (WCSPH) method, we show that the flow classification of tidal bores must explicitly take into account the frame of reference. When the river flow does not reverse after the bore passing, this is a consequence of the perspective imposed by the background current (itself resulting from the previous tide and the natural river discharge toward the sea). Depending on the relative magnitudes of the river current and the velocity field induced by the bore, one may observe a global reversal, a local reversal (or both), or no reversal at all. When the tide is spatially inhomogeneous, with a tidal range that varies along the estuary rather than a simple step in water depth, transitions between bore types may occur during the bore passing.
\end{abstract}

\maketitle

\section{Introduction}
\begin{figure}[!htbp]
\centering
\includegraphics[width=0.35\textwidth]{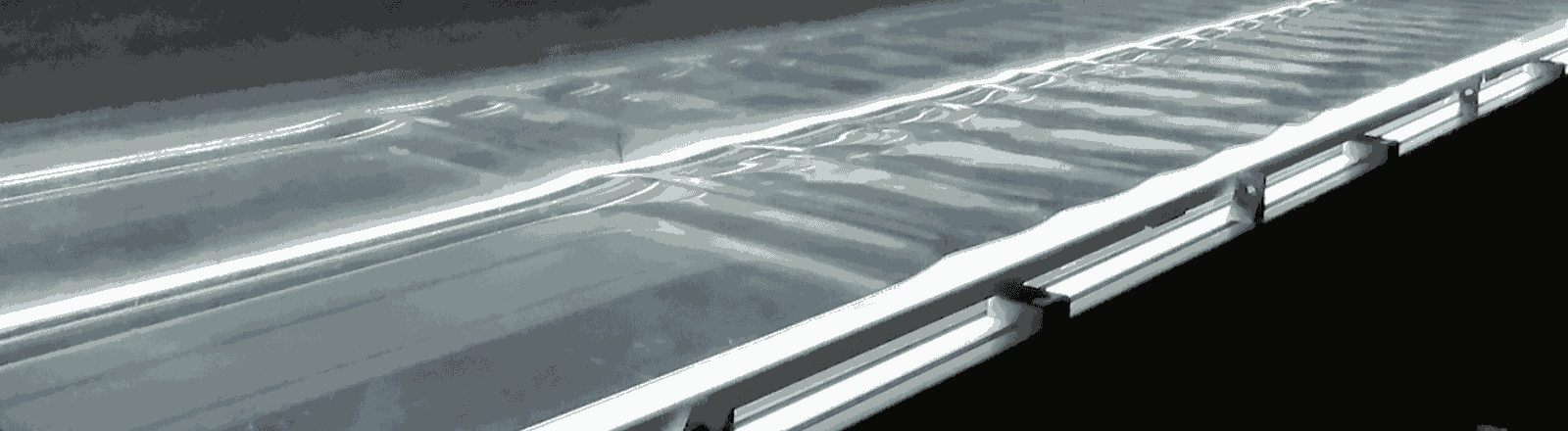}
\includegraphics[width=0.35\textwidth]{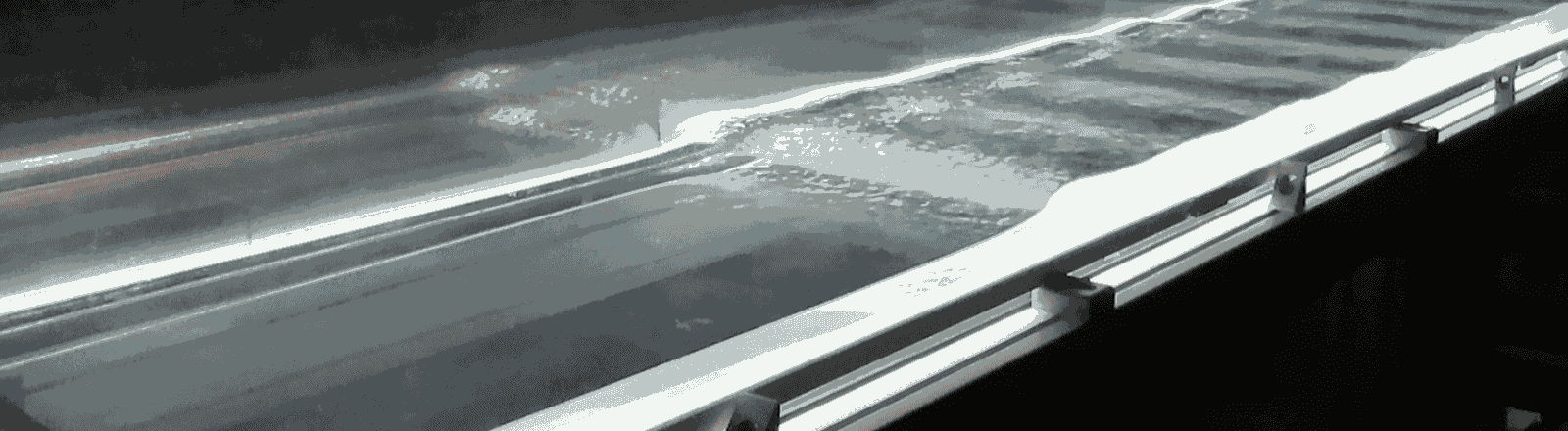}
\includegraphics[width=0.35\textwidth]{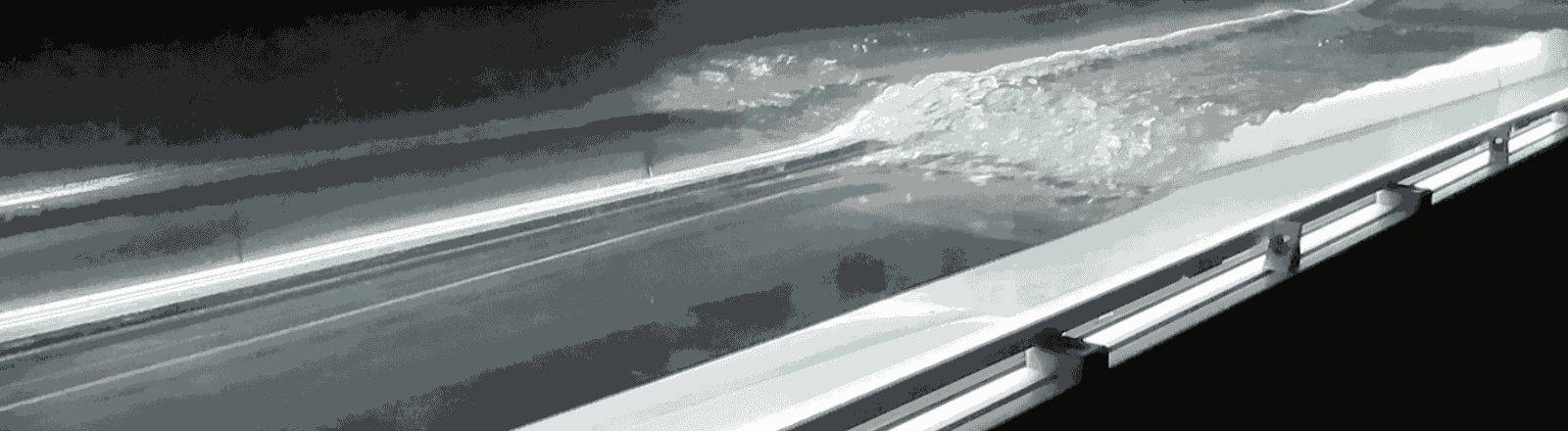}
\caption{Experimental visualizations of laboratory analogue tidal bores with a LED lighting system obtained by the volume inversion method of Rousseaux {\it et al.} (2016) \cite{rousseaux2016novel} (top) undular bore ($h_s=31.8\,mm$ and $q=0.0344\,m^2/s$); (middle) undular breaking bore ($h_s=55\,mm$ and $q=0.0346\,m^2/s$); (bottom) breaking undular bore ($h_s=105\,mm$ and $q=0.0647\,m^2/s$). The width of the rectangular open water channel is $W=0.385\,m$. $h_s$ is the initial static depth and $q$ is the flow rate per unit width before turning off the pump.}
\label{TBclass1}
\end{figure}

A tidal bore is a nonlinear surge that propagates upstream of estuaries with an important tidal range usually at low water level \cite{sene2021tidal, roy2023tidal}. Since the seminal work of Henry Favre \cite{favre1935etude}, four types of surges have been distinguished depending on whether they propagate upstream or downstream and whether they are positive or negative. The tidal bore corresponds to a positive surge propagating from downstream against the river flow. This is observed for example in the Severn River (United Kingdom) \cite{arildsen2025tidal, sene2025estuarine}, the latter river current being the addition of the flow current created by the previous tide and the nominal flow of the river towards the sea directed by the natural slope of the river thalweg. In some estuaries, such as the Mont Saint Michel Bay (France), the bore may even form in the absence of river discharge and propagate like a dam-break wave over the exposed sediment bed before entering the Sée River \cite{furgerot2016, cosma2022sedimentology}. Several bore types have been reported based on free surface observations \cite{dolgopolova2013conditions, dolgopolova2017tidal} (see Fig. \ref{TBclass1}): undular, partially breaking, or fully breaking, with or without local or global flow reversal at the passage of the front. 
This highlights the crucial role of the river current in shaping bore dynamics. Equally important is the frame of the observer, which determines how the relative motion between the bore and the river flow is perceived: compare the testimonies of a surfer with an observing scientist.\\

In this work, we propose to test a bulk classification that goes beyond the naive one (based on the visual inspection of the free surface) depending on whether the river current or the tidal bore wins the battle. Indeed, any flow classification assumes an underlying reference frame. All the observations of tidal bores were made so far in the reference frame of the river banks 
since it is the natural frame where the observer set the measurements system like for instance a scale to measure the water depth with its modern exemplification with the use of observation tower and photogrammetry technique \cite{huang2021experimental}. The scale sensor is a universal one since all the observers will agree on the same water depth measured. The speed, being a relative quantity, depends on the frame of reference. Water depth is a Galilean invariant quantity, whilte velocity is not. Two alternative viewpoints are therefore relevant in addition to the river bank frame  
: the frame of the river current 
and the frame of the bore front.
A central question then arises: how can we compare the descriptions of bore dynamics obtained in these three frames, and which one is most appropriate for classification? To address this question, we combine PIV (Particle Image Velocimetry) measurements with classical laboratory methods for generating tidal bores \cite{thomas2022eulerian, rousseaux2016novel} and  numerical simulations using WCSPH (Weakly Compressible Smoothed Particle Hydrodynamics) method  \cite{lucy1977, monaghan1992smoothed}. This Lagrangian approach allows us to observe all theoretically predicted regimes in the three frames and to compare them systematically.  

Our scope is restricted to hydrodynamics under the simplifying assumption of a uniform plug flow. Boundary layer effects and turbulence, recently discussed in \cite{carneiro2024hydrodynamic, lin2025velocity} are not included. {The interaction between the bore and bathymetry is also beyond the present scope, although it is known to be essential. As the french engineer Henri Partiot noted \cite{partiot58, partiot61}, \textit{"when the tide comes in the bay, a simple change in the direction of the currents shows that the sea gets into the deeper passes. On the banks of those where the depth is less, we see a wave whose one of the ends travels along the bank by breaking and extends but without breaking towards the middle of the pass: this wave is nothing but the tidal bore. In shallow areas or on the banks, it is a long wave that breaks and travels like a white line across the bay; this is a water roller announcing the flood and who is its head, so to speak. It is sometimes followed by several waves, the firsts are the only foaming, and decrease successively in height. These waves are known on the Seine to as stubbles. In the deeper parts where the tidal bore is formed by a shoal located downstream, it spreads like an undulation. It is always followed by a sudden elevation of the water level. When it occurs on a bank and is in the form of a water roller, it is the edge of the water layer that travels"}. 
The formation and hydrodynamics of the tidal bore are sensitive to the general morphology of funnel shaped estuaries. They are not yet fully understood due to their spatial and temporal variability \cite{violeau2022, jouy2024-1, jouy2024-2}. The development and characteristics of the tidal bore depends on the spatial evolution of tidal range in estuaries, the shape of the tidal wave (tidal asymmetry), the intensity of river flow, anthropization and particularly to local bathymetric variations \cite{zhang2024structures, santos2025dynamics}. 

Short term bathymetric variations are mainly induced by erosion and deposition cycles partly controlled by the tidal bore itself, or periods of high or low water. Investigations are needed to characterize the factors controlling the tidal bore development in estuaries whose bathymetry changes due to the tidal bore induced erosion. The resulting sediment transport will also not be taken into account in this work. The hydro-sedimentary processes associated with the tidal bore is still an unresolved issue  \cite{furgerot2016, reungoat2017, tessier2017}. For example, the question of the feedback of the sediment on the flow profile hence on the classification is an open question \cite{dolgopolova2020mechanism, qiu2025hydrodynamics}.

The experimental and numerical classification discussed usually in the literature is the one relying on the observation of the free surface \cite{pan2026} (see Fig. \ref{TBclass1}). Mathematically, it relies on the value of a non-dimensional number, the Kelland number $K_e=\frac{V_b-(-V_1)}{c_m}=\frac{V_b+V_1}{\sqrt{gA_r/W}}$ \cite{kelland1840xxvi, chanson2013environmental} that encompasses the effect of the nominal river current and which compares the speed (in the reference frame of the river bottom or banks)
of the bore front $V_b$ Doppler-shifted by the river counter-current speed $V_1$ to the speed $c_m$ of a linear long gravity wave in shallow water at rest (with $c_m=\sqrt{gh_1}$ where $h_1=A_r/W$ is the equivalent water depth namely the ratio between the river vertical area section $A_r$ to the mirror width of the water channel $W$). Say differently, it could be viewed as the ratio between the front speed in the river frame of reference ($V_b-(-V_1)=V_b+V_1$) to the dispersionless linear water wave speed $c_m$ in the water channel specific geometry. When the latter is rectangular with water depth $h=h_1=Wh/W$, the less-known Kelland number becomes the well known Doppler-shifted height Froude number $F_r=(V_b+V_1)/\sqrt{gh}$. 

The Froude number $F_r$ for a rectangular geometry is the canonical non-dimensional number employed in laboratory experiments with rectangular open channels. In \cite{chanson2013}, Chanson showed that the Kelland number $K_e$ introduced itself naturally when combining non-dimensional flow and momentum conservation equations. The transition from one free surface deformation to the other happens at definite threshold values (with values such that $K_e>1$) whose origin are still unexplained. Another non-dimensional parameter discussed in the literature is the ratio between the upstream and downstream depths of the tidal bore front that may be related to the Froude number in peculiar situations like in the hydrostatics approximation \cite{putra20192d}. The river Reynolds number ($R_e=V_1h_1/\nu$ where $\nu$ is the water kinematic viscosity) is usually very large ($10^4-10^6$) since the natural and laboratory flows are turbulent. The river speed is usually measured with punctual sensors like ADV (Accoustic Doppler Velocimetry) \cite{furgerot2016, reungoat2017, tessier2017, chanson2024acoustic} and the bore front speed can now be accessed through the use of drones for camera tracking \cite{zhang2024automatic}.

Another theoretical and numerical flow's classification was proposed recently \cite{berchet2018flow} based on the observation of the velocity field beneath the free surface in the frame of reference of the banks
that we test here (see Fig. \ref{TBclass2}). In the work by Berchet $et \, al.$ (2018) \cite{berchet2018flow}, the reversal of the current flow was observed in their numerical simulations on undular tidal bores in an open channel. They distinguished, by selecting different initial conditions, using two numerical methods for generating tidal bores, four types of undular tidal bores that exhibit an inversion of the flow direction or not during the bore passing, despite similar Froude numbers, $1.1 < F_r < 1.2$.

Field observations \cite{wolanski2004undular, reungoat2014sediment, bonneton2015tidal, furgerot2016sediment} and laboratory experiments \cite{rousseaux2016novel, thomas2022eulerian} have confirmed the existence of these regimes, although real estuaries often exhibit more complex behaviour due to transverse variations and tidal inhomogeneity. In this work, we show that this classification is incomplete because it implicitly assume a specific frame. By analysing the flow in three separate frames of reference, we propose a unified classification that distinguishes intrinsic bore type (undular vs breaking) from reversal behaviour (none, local, global).

\begin{figure}
\centering
\includegraphics[width=0.4\textwidth]{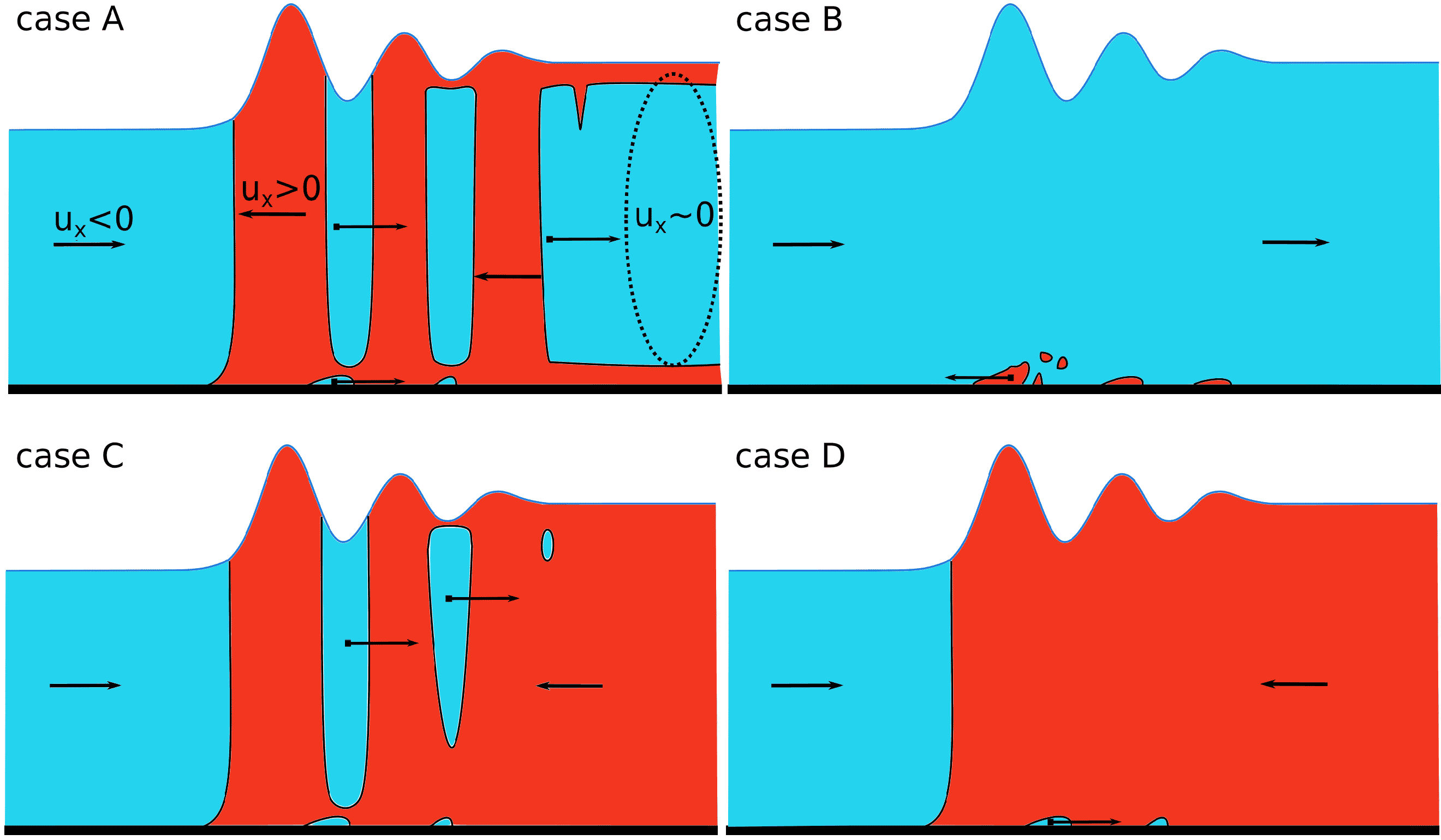} 
\caption{Simplified schematics, shown in the riverbank reference frame
,  of the longitudinal velocity component $V_x$ during the development of an undular bore, with emphasis on the bore front and its first stubbles or whelps (secondary waves). Red regions indicate downstream directed flow $V_x = V_1 > 0$, aqua shade blue regions indicate upstream directed flow $V_x = V_1 < 0$, and black contours correspond to the zero velocity isocline $V_x = 0$, following the classification of Berchet et al. (2018) \cite{berchet2018flow}.}
\label{TBclass2}
\end{figure}

Cases A and B were generated by the impact of the flow against a completely or partially closed vertical gate. As for the other two cases, C and D, they were generated using two rectangles of water with two different heights and flow velocities: 
\begin{itemize}
\item Case A, hydrodynamics dominated by undulations, started with a gate completely closed, $h_g = 0\, m$. Before the wavefront, the horizontal component $V_x$ of the flow velocity was completely negative. Under the waves train, the longitudinal component $V_x$ of the flow velocity flowed alternatively upstream in the troughs and downstream in the crests with a range of the velocity magnitude progressively decreasing. 
\item Case B, hydrodynamics dominated by the initial current, started with a gate partially closed and a gate opening of $h_g = 0.1\,m$. The longitudinal component $V_x$ of the flow velocity decelerated and re-accelerated beneath the wave crest and wave trough, respectively, contrarily to Case A where it changes direction. No reversal of the longitudinal component $V_x$ of the flow velocity occurs in the water column.  
\item Case C, hydrodynamics partially dominated by undulations and then bore flow, started by the collapse of the higher water domain, $h_2 > h_1$ at $x = 0\, m$, over the initial flow. Beneath the bore front and the first undulations, the flow was similar to Case A. Beneath the secondary undulations, the surface of areas where the horizontal component $V_x$ of the flow velocity was negative beneath the trough progressively diminished until the areas disappeared. In the following secondary undulations, the hydrodynamics of the flow were similar to those in Case D.
\item Case D, hydrodynamics dominated by the bore flow, started with the collapse of the higher water domain, $h_2 > h_1$ at $x = 0\,m$, over the initial flow. During the bore passing, the horizontal component $V_x$ of the flow velocity was larger beneath the crest than beneath the trough. Beneath crests, the velocity was more intense both near the free surface and near the bed than in the rest of the water column. For the following secondary undulations, the longitudinal component $V_x$ of the flow velocity followed a similar pattern to the one observed beneath the wave front and trough. For this case, the flow suddenly changed direction beneath the bore front and undulated regularly after the bore front. 
\end{itemize}

This bulk classification based on the velocity pattern beneath the free surface assumes a constant tidal range hence no inhomogeneity in time of the tide which is clearly an idealisation as well as the absence of transverse variation of the velocity profile due to say the bathymetry that induces also a transverse variation of the speed of the nominal current in the river. Studies in the field have already shown more complicated situations: indeed, still according to Partiot \cite{partiot58, partiot61}.

A rapid review of some recent field or laboratory works show the existence of some regimes compatible with the theoretical predictions:
\begin{itemize}
\item Wolanski {\it et al.} in 2004 \cite{wolanski2004undular}: Dale River (Australia) with an undular bore and a global flow reversal at front passage followed by speed oscillations with the same sign = Case D.
\item Reungoat {\it et al.} in 2014 \cite{reungoat2014sediment}: Garonne river site at Arcins channel (France) with an undular bore and a global flow reversal at front passage followed by speed oscillations with the same sign = Case D.
\item Bonneton {\it et al.} in 2015 \cite{bonneton2015tidal}: Garonne river site at Podensac (France) with an undular bore and a global flow reversal at front passage followed by speed oscillations with the same sign = Case D.
\item Furgerot {\it et al.} in 2016 \cite{furgerot2016sediment}: Sée river site at Le Bateau (France) with an undular bore and a flow reversal invading the water column from the bottom at front passage followed by speed oscillations with the same sign  = Case $D_{delay}$.
\item Rousseaux {\it et al.} in 2016 \cite{rousseaux2016novel}: Institute Pprime laboratory experiments using the volume inversion method of an undular bore with local and global flow reversals at front passage with the elephant foot pattern = Case C.
\item Thomas and David in 2022 \cite{thomas2022eulerian}: Institute Pprime laboratory experiments using the partially closed-gate method with an undular-breaking bore without local or global flow reversals = Case B.
\end{itemize}
Of course, in the case of global flow reversal with or without a delay at the front passage in the field works, there is always a latter global reversal at the end of the tide but which is uncorrelated from the tidal bore, the feet of the tide somehow.

The tidal bores are classified in the family of free surface flows in which the complex interface configurations can be involved \cite{chanson2010, chanson2024}. The fluid dynamics of free surface flows implie that carrying out a numerical study of tidal bores remains a challenge for all numerical methods \cite{simon2023}. In the SPH (Smoothed Particle Hydrodynamics) methods, the governing partial differential equations are approximated with the kernel approximation in continuous form, and the particle approximation in discretization form \cite{rahmati2025}. The accuracy of two approximations depends on the distribution of fluid particles. If the distribution of fluid particles is not uniform, the properties of the kernel function are not checked in the particle approximation. Thus a discrepancy between kernel approximation and particle approximation appears \cite{liu2022}. In the bore passing, the flow disrupts the distribution of fluid particles at the free surface. The SPH methods can simulate negative pressures. The presence of negative pressures can cause particles to cluster due to loss of repelling forces. In the SPH community, this problem is called the particle clustering \cite{ghoneim2025}. 

To overcome this, two approaches have been developed and continue to be developed. The first family consists in regularizing the particle distribution. In this family, there are the remeshing method \cite{letouze2013}, the artificial viscosity method \cite{monaghan1983}, the particle shifting method \cite{monaghan1983}, the particle collision method \cite{kruisbrink2018} and the background pressure method \cite{morris1997}. The calibration of these different methods is difficult \cite{bagheri2024, fan2024}. The SPH community knows that the choice of the kernel function and the smoothing length parameter directly impacts the stability and accuracy of SPH methods \cite{wang2024, li2024}. This is the second approach. In this approach, the Cubic spline and Wendland kernel functions are analyzed \cite{monaghan1985, wendland1995}. The particle clustering is often more noticeable with the Cubic spline kernel function \cite{dehnen2012}. Works are currently underway to improve these two kernel functions \cite{liu2010}.    

The first attempts to numerically study the tidal bores date to works by Furuyama and Chanson in 2008 \cite{furuyama2008}. In their works, Furuyama and Chanson simulated tidal bores with a numerical model based on the CIP-CUP (Cubic Interpolated Pseudo-particle Combined Unified Procedure) method equipped with a LES (Large Eddy Simulation) model and a re-initialization method, proposed by Yabe and Wang in 1991 \cite{yabe1991}. Resolving the advection phase of governing equation with the CIP (Cubic Interpolated Pseudo-particle) method, the CIP-CUP method can be defined as a semi-Lagrangian method \cite{yabe1991bis}. In the Lagrangian methods represented by the SPH (Smoothed Particle Hydrodynamics) methods, the setting of the free surface boundary condition is easier than with the Eulerian methods, as the VOF (Volume Of Fluid) and Level-Set methods \cite{monaghan1994, bonet1999, landrini2007, liu2015, nikeghbali2018, nikeghbali2024}. Furuyama and Chanson prefered to use the CIP-CUP method instead of a SPH method because these methods need a large number of particles for good accuracy and a search process to find the neighbor, inducing a high CPU (Central Processing Unit) time \cite{furuyama2008}.

To reduce the CPU time, the first implementation of SPH methods totally on GPU (Graphical Processing Unit) using the CUDA (Compute Unified Device Architecture) language was realized by Kolb and Cuntz in 2005 \cite{kolb2005}, and Harada {\it et al.} in 2007 \cite{harada2007}. In the SPH methods introduced by Lucy in 1977 \cite{lucy1977}, Gingold and Monaghan in 1977 \cite{gingold1977}, there are two families of methods. The ISPH (Incompressible Smoothed Particle Hydrodynamics) methods solve the Laplace equation to estimate the pressure \cite{ramos2020}. As for the WCSPH (Weakly Compressible Smoothed Particle  Hydrodynamics) methods, the pressure is estimated using an equation of state (trait equation of state). The GPU implementation is computationally efficient, especially in the WCSPH approach \cite{herault2010}. The use of implementation of a WCSPH method totally on GPU gives DualSPHysics, an open source code, the advantage of being very performant with fast numerical simulations \cite{crespo2015, luo2021, letouze2025}. DualSPHysics is a hardware accelerated SPH code developed based on FORTRAN SPH formulation implemented in SPHysics \cite{gomez-gesteira2004, zhan2025}. The code was developed to solve free surface flow problems requiring high computational cost \cite{dominguez2022, heydemans2024}. 

In the scientific literature, the first 2D (Two Dimensions) numerical simulations of tidal bore applications with the SPH methods have been performed by Monaghan in 1994 \cite{monaghan1994}, Bonet and Lok in 1999 \cite{bonet1999}, and Landrini {\it et al.} in 2007 \cite{landrini2007}. As indicated by Lui {\it et al.} in 2015 \cite{liu2015}, these studies tended to use the tidal bore only as one benchmark test. Recently, the SPH methods have been used to study the bore passing through an open channel with a complex geometry by performing 3D (Three Dimensions) numerical simulations with the software DualSPHysics \cite{gomez-gesteira2010, gomez-gesteira2012}. In 2010, Nouri {\it et al.} investigated numerically the characteristics of tidal bores at the impact point with a square column in an open channel \cite{nouri2010}. In 2012, St-Germain {\it et al.} simulated numerically in 3D the impact with columns of tsunami waves rapidly flowing on wet and dry beds. The resulting pressure versus time and force versus time on a structure and wall positioned vertically due to the bores impact were compared quantitatively \cite{stgermain2014}. In 2013, De Padova {\it et al.} showed that the 3D numerical simulations were required to study undular tidal bores in a very wide open channel because of effects of oblique shock waves generated by the presence of lateral walls \cite{depadova2013}. In 2015, Liu {\it et al.} investigated numerically the effects of a short channel constriction imposed with bridge piers in an open channel on an undular tidal bore \cite{liu2015}. 

However, in the case where the open channel presents a simple geometry, no constriction or obstacle, the tidal bore remains 2D. Thus the 2D numerical simulations are largely sufficient to study the characteristic of tidal bore in this configuration of open channel with DualSPHysics. In 2012, Federico {\it et al.} performed 2D numerical simulations of several types of jumps to investigate the effects of different upstream and downstream conditions with specific focus on the velocity field, pressure forces, water depths and location of the jump \cite{federico2012}. In 2017, De Padova {\it et al.} performed 2D numerical simulations of the transition from supercritical to subcritical flow at an abrupt drop, which can be characterised by the occurrence of oscillatory flow conditions between two different jump types. The physical quantities, free surface profile downstream of the jump, velocity, pressure and vorticity, were studied. In the SHP model, an algebraic mixing-length model and a two-equation model were used to represent turbulent stresses \cite{depadova2017}. In 2018, Nikeghabli {\it et al.} showed numerically the effects of breaking and undular bores on a movable bed. The numerical simulations were 2D with a two phase model implemented in the software DualSPHysics \cite{nikeghbali2018}. In 2019, Lin {\it et al.} simulated numerically the 2D hydraulic jump in open channel with different inflow conditions \cite{lin2019}. In 2025, Zheng {\it et al.} performed 2D numerical simulations to quantify hydrodynamic characteristics of negative and undular surges in an impounded channel. For undular surges, empirical correlations quantified propagating wave parameter variability, while vertical velocity profiles displayed linear stratification across the water column, validating Boussinesq’s hypothesis \cite{zheng2025}. 

In the following, the 2D numerical simulations performed with DualSPHysics will be compared to the experimental data measured in the rectangular open channel of the Institute Pprime, in which no constriction or obstacle is fixed. Presenting a rectangular transverse section, the open channel is $L= 7\,m$ long with a wide of $W=0.385\,m$. The experiments were done with either the pump stop method, the partially or fully closed gate method (see Fig. \ref{bore-setup}). The measurements were obtained with the PIV technique (see Fig. \ref{TBPIV}). It is the most commonly used technique in the laboratory studies \cite{thomas2022} for capturing the velocity fields (see Fig. \ref{experdata}). It is included  experiments with three methods of generation of tidal bores: the closing gate method, the opening gate method, and the pump stopping method. A comparison with one of the experiments with the closing gate method will be done at first to first validate the numerical model. The comparison will be done using the spatial distribution of the horizontal components $V_x$ of the flow velocity, the free surface (although the PIV technique is not a method developed for free surface), and the streamlines useful for the observations of some structures as the elephant foot pattern characterization in case of local river reversal \cite{rousseaux2016}. The experimental data will also be used later to present the cases which were observed as a qualitative comparison with the numerical simulations.

\begin{figure}
\includegraphics[scale=0.3]{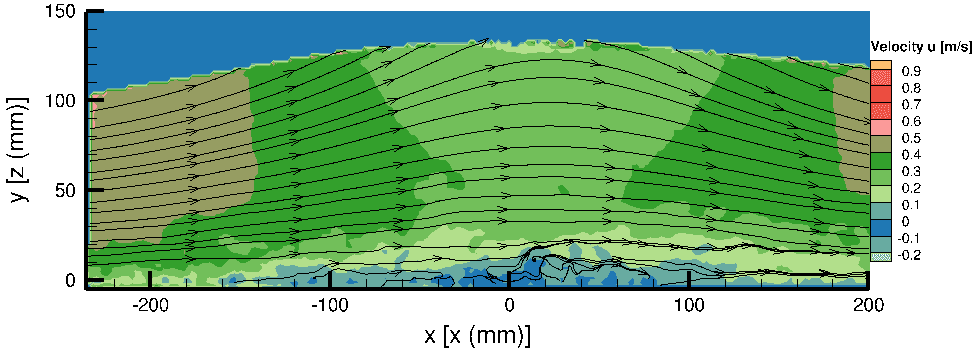} 
\includegraphics[scale=0.3]{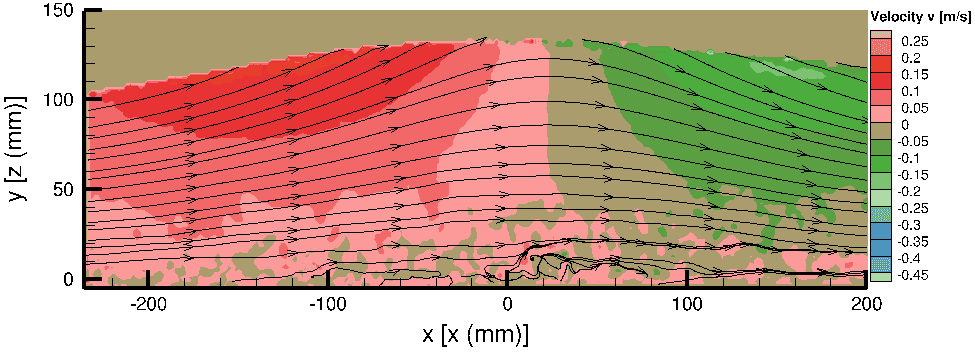}  
	\caption{Spatial distribution of the two velocity components obtained from the PIV measurements: horizontal velocity $V_x$ (top) and vertical velocity $V_z$ (bottom). Snapshot of the experiment representing a transition case: case B at early stage (positive velocity vectors) turning into case A. The case is obtained using the partially closed gate method with $V_1=0.61\, m.s^{-1}$, $h_1=0.094\,m$, $Fr = 1.25$ and $Re=57, 340$ and is classified as case B in the current snapshot. The case is classified as case A later in the paper (see Fig. \ref{fig:figuresoil1} - bottom).  Streamlines, computed from the instantaneous velocity field, are superimposed and colour coded by the local speed magnitude. Data visualized using Tecplot.}
\label{experdata} 
\end{figure}

\section{Generation of a tidal bore}

\subsection{Numerical model}

The $2D$ numerical simulations are done using the configuration in Figure \ref{fig:geometry}. The geometry of computational domain is generated using FreeCAD [\url{https://www.freecad.org/}], the numerical simulation of the tidal bores are then done using DualSPHysics [\url{https://dual.sphysics.org/}].

\begin{figure}
\includegraphics[scale=0.5]{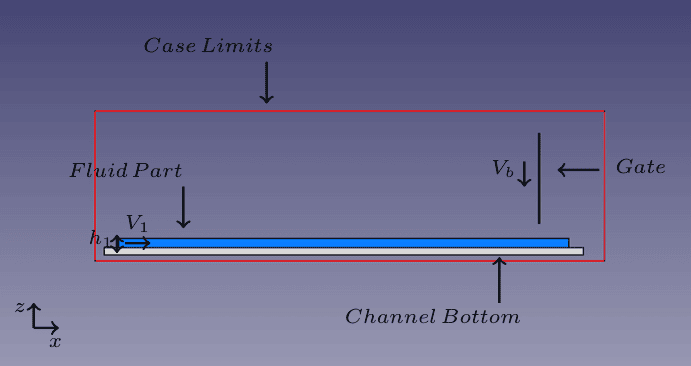}
\caption{2D computational domain to generate a tidal bore using the closing gate method for the WCSPH simulations. The geometry is created in FreeCAD and discretized into SPH particles presenting either fluid or solid boundaries (channel bottom and gate). The gate falls at a prescribed closing speed $V_g$, leaving a residual opening height $h_g$. Fluid particles carry out the standard SPH physical properties with an evolution following the WCSPH formulation described in Eqs. (1)–(10).}
\label{fig:geometry}
\end{figure}

%
%
%
%
%
%
%

The computational domain $\Omega$ is discretized into particles, each holding physical properties. In the SPH formulation, a physical property $f$ is calculated using a convolution product  \cite{lucy1977} \cite{monaghan1992smoothed}:
\begin{equation}
f(\textbf{r}_i) = \int _{\Omega} f(\textbf{r})\textit{W}(\textbf{r}_i-\textbf{r},h)
d\Omega
\label{eqn:conv}
\end{equation} 
where $\textbf{r}_i$ is the position of the particle $i$ and $W$ is a smoothing kernel function with $h$ being its smoothing parameter. By writing $m_{j} = \Delta V_{j} \rho_{j}$ the mass of the particle $j$ with $\rho_j$ its density and $\Delta V_j$ its elementary volume ($\Delta V_j = dp^2$ with $dp$ the particle size; the distance between the particles), the equation (\ref{eqn:conv}) can be rewritten in the discrete form for a number of particles $N$ as:
\begin{equation}
f(\textbf{r}_i)\approx\sum _{j=1}^N \frac{m_j}{\rho _j}f(\textbf{r}_j)W_{ij} \quad \mbox{with} \quad W_{ij} = W(\textbf{r}_i-\textbf{r}_j,h).
\label{interpolation}
\end{equation}
In DualSPHysics, two kernel functions are implemented; Cubic B-spline \cite{monaghan1985} and Wendland \cite{wendland1995} functions. Their expressions are given for $q = \frac{|\textbf{r}_{ij}|}{h}$ as follows: 
\begin{equation}
W_{W} (q)= \frac{\alpha_{d}}{h^d} 
\left \{
\begin{array}{rcl}
\left(1-\frac{q}{2}\right) ^{4}(2q+1) &\quad & 0\leq q \leq 2, \\
\hspace*{-2cm}0 &\quad & \; \text{otherwise.}
\end{array}
\right .
\label{wendland}
\end{equation}
\begin{equation}
W_{C}(q)= \frac{\alpha_{d}}{h^d} 
\left \{
\begin{array}{rcl}
 1-\frac{3}{2}q^{2}+\frac{3}{4}q^{3}&\quad & \; 0\leq q \leq 1 , \\
\hspace*{-1cm}\frac{1}{4}(2-q)^{3}&\quad & \; 1\leq q \leq 2, \\
\hspace*{-2cm}0 &\quad & \; \text{otherwise}.
\end{array}
\right.
\label{cubic}
\end{equation}
where $W$ and $C$ are the indexes indicating the type of the kernel function, $W$ = Wendland function and $C$ = Cubic B-spline function, $\textbf{r}_{ij} = \textbf{r}_{i} - \textbf{r}_{j}$ is the gap between the location of two particles, $\alpha_{d}$ is the constant of normalization with the value of $\frac{3}{2}$ in $1D$ (One Dimension), $\frac{7}{4\pi}$ in $2D$ (Two Dimensions) and $\frac{21}{16\pi}$ in $3D$ (Three Dimensions) for the Wendland function, and $\frac{2}{3}$ in $1D$, $\frac{10}{7\pi}$ in $2D$ and $\frac{1}{\pi}$ in $3D$ for the Cubic B-spline function. During the test of the kernel functions implemented in DualSPHysics, a validation of the model using the experimental data showed that the Wendland kernel performed best. The Wendland function is thus used here with a smoothing parameter $h = 0.004$ and a particle size $dp = 0.002\,m$ (distance between particles). The ratio $h/dp$ is thus fixed to 2, as mostly used in the SPH methods \cite{jandaghian2022}. The comparative study between the two kernel functions is presented in the supplement material of this work A.  

The compressible Navier-Stokes continuity and momentum equations, used to describe the fluid flow, can be written in SPH Lagrangian form as follows \cite{CRESPO2015204}:
\begin{equation}
\frac{d\rho _i}{dt}= \rho_i \sum _j \frac{m_j}{\rho_j} \textbf{V}_{ij}\cdot\nabla W_{ij}
\quad \mbox{with} \quad \textbf{V}_{ij} = \textbf{V}_{i} - \textbf{V}_{j},
\label{eq:densite}
\end{equation}

\begin{align}
\frac{d \textbf{V}_i}{dt} &=
-\sum _j m_j\left( \frac{P_j}{\rho _j^2} + \frac{P_i}{\rho _i^2} \right) \nabla _iW_{ij} + \textbf{\boldmath$g$} \\ \nonumber
&+ \, \sum _jm_j\left( \frac{4\nu_0 \textbf{r}_{ij}\cdot \nabla _iW_{ij}}{(\rho _i+\rho _j)(r_{ij}^2+\eta ^2)}\right) \textbf{V}_{ij} \\ \nonumber
&+ \, \sum _{j} m_{j} \left( \frac{\textbf{\boldmath$\tau$ }^{j}}{\rho _{j}^{2}} +\frac{\textbf{\boldmath$\tau$ }^{i}}{\rho _{i}^2} \right) \nabla _{i}W_{ij}
\quad 
\label{momentum}
\end{align}
where $\eta^2 = 0.01 h^2$, $\textbf{V}_{i}$, $P_i$ and $\textbf{\boldmath$\tau$ }^{i}$ are the fluid velocity, water pressure and stress tensor of the particle $i$. $\textbf{\boldmath$g$}$ is the gravity vector. $\nu_0$ = $10^{-6}\,m^2/s$ is the water viscosity. Since we are dealing with high Reynolds numbers $Re > 4,000$, the viscosity term is divided into two part: the laminar viscosity represented in the third term of the momentum equation (6) \cite{monaghan1992smoothed}, and a SPS (Sub-Particle Scale) stress tensor expressing the turbulence in the last term of the equation (6) \cite{dalrymple2006numerical}. The stress tensor $\textbf{\boldmath$\tau$ }^{i}$ of the particle $i$ is given by:
\begin{equation}
\tau^i_{ab} = \rho_i \nu _t \left( 2 S^i _{ab}-\frac{2}{3}k\delta _{ab} \right) - \frac{2}{3} 0.0066 \, \Delta l^2 \delta _{ab} |\textbf{\boldmath$S$}^i|^2.
\end{equation}
The subscript $ab$ indicates the spatial directions $a$ and $b$ for the components of the stress tensor $\textbf{\boldmath$\tau$ }^{i}$. $\nu _t = (C_S \Delta l)^2|\textbf{\boldmath$S$}^i|$ is the turbulent eddy viscosity. $C_S = 0.12$ is the Smagorinsky constant. $\Delta l$ is the particle to particle spacing and $\textbf{\boldmath$S$}^i$ is the SPS strain tensor of the particle $i$. $k$ is the SPS turbulence kinetic energy. $\delta_{ab}$ is the Dirac distribution between the spatial directions $a$ and $b$. In the SPH formulation \cite{CRESPO2015204}, each particle $i$ is moved with a velocity close to the average of particle velocities in its neighbourhood using:
\begin{equation}
\frac{d\textbf{r}_i}{dt}=\textbf{V}_i + \varepsilon \sum _j \frac{m_j}{\overline{\rho _{ij}}}\textbf{V}_{ij} W_{ij}
\quad \mbox{with} \quad {\overline{\rho _{ij}}} = \frac{\rho_i+\rho_j}{2}.
\label{eq:moving}
\end{equation}
$\varepsilon$ is a constant of the problem ($0\leq \varepsilon \leq 1$). The equation (\ref{eq:moving}) uses a corrective term so that close particles move with approximately the same velocity.
In the WCSPH method, the fluid is treated as weakly compressible. An equation of state is used in order to determine the fluid pressure $P$ based on its density $\rho$ \cite{monaghan1994}\cite{monaghan1999gravity}\cite{moffatt2002george}:
\begin{equation}
P = B\left[ \left(\frac{\rho}{\rho _0}\right)^{\gamma}-1\right]
\quad \mbox{with} \quad B = \frac{c_o^2 \rho_0}{\gamma}.
\label{state}
\end{equation}
$\gamma = 7$ is the polytropic index of water, $\rho _{0} = 1000\,kg/m^{3}$ is the density of reference and $c_0 = \sqrt{\frac{\partial P}{\partial \rho}}\vert _{\rho _0}$ is the speed of sound at the density of reference. The equation (\ref{state}) gives a direct relation between the pressure in the fluid and the local density, an advantage of the weakly compressible SPH method.

\begin{figure}
\includegraphics[width=0.4\textwidth]{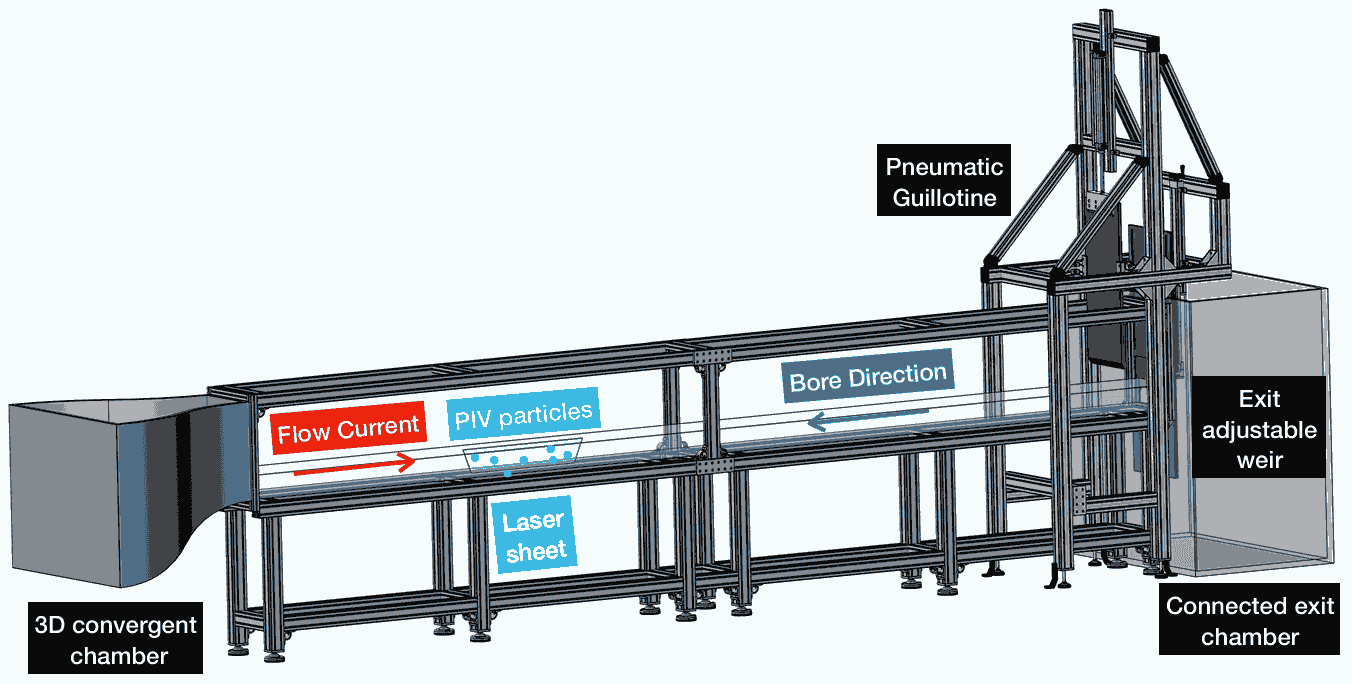} 
\caption{Schematic view of the experimental flume setup, showing the flow direction and the upstream propagating bore. The facility consists of a 3D convergent inlet chamber, an open channel test section, and a connected downstream chamber with an adjustable weir. A pneumatically actuated vertical guillotine gate may be used to generate the tidal bore. The location of the laser sheet and PIV measurement region is also indicated.} 
\label{bore-setup}
\end{figure}

The free surface flow is governed by the compressible Navier-Stokes equations to which the initial and boundary conditions are added to ensure the existence and uniqueness of the solution. The modified Dynamic Boundary Condition (mDBC) implemented in DualSPHysics is used to avoid the limitations of the DBC boundary conditions \cite{CRESPO2015204}. The (.xml) file is generated using FreeCAD where the  geometry is defined and the parameters before the numerical simulations are run using DualSPHysics. The closing gate method is used in our numerical simulations generated using boxes in FreeCAD. The blue part represents the fluid part. As for the channel bottom and gate, they are represented in gray, and are treated as solids (see Fig. \ref{fig:geometry}). The gate goes down with a closing speed $V_g$ leaving an opening $h_g$. Since we have a solid part, channel bottom and gate, we also have the no-penetration and no-slip conditions of the fluid adhesion to the wall, given as following:
\begin{equation}
\textbf{V}.\textbf{n}=0 
\quad \mbox{and} \quad
\textbf{V}.\textbf{t}=0
\label{eq:tangente}
\end{equation} 
where $\textbf{n}$ and $\textbf{t}$ are the normal and tangent vectors of solid boundaries, respectively. 

The previous equations are solved using the second order Simplectic Position Verlet Scheme, an adapted scheme for Lagrangian methods as it is a time reversible scheme \cite{CRESPO2015204}. The time step $\Delta t$ is variable during the numerical simulations. It is related to the speed of sound $c$ and is dependent on the Courant Friedrich Levy (CFL) condition to ensure the convergence and stability of numerical scheme \cite{CRESPO2015204}.

\subsection{Experimental setup}

Experiments were performed at the Institute Pprime (Futuroscope, France) in the Environmental Hydrodynamic Platform’s open water channel SedHyde (see Fig.~\ref{bore-setup}). The channel is $7\,\mathrm{m}$ long, $0.385\,\mathrm{m}$ wide, and $0.6\,\mathrm{m}$ high. It features a PCM (Pompes Compresseurs Mécaniques) Moineau pump with an Archimedes screw controlled with a Schneider Electric variable frequency drive with a maximum flow rate of $Q=65 \, L/s$ as measured with an Endress + Hauser Promag 55 electromagnetic flow-meter. At the channel inlet, a 3D convergent chamber with two grids with different mesh sizes and a honeycomb ensure a plug like flow with small fluctuations around its average with a tiny boundary layer. 

The dynamical water level depends on the entrance flow rate and its initial static level ($h_i=11.8\,cm$ here). It may be adjusted in addition by a downstream gate at the connection between the main flume and the exit chamber that are geometrically linked. The difficulty of maintaining a constant downstream deph in the channel allows to model the in situ conditions where the bores vary both temporally and spatially within estuaries. One example is the Qiantang estuary. In their works, Huang et al. used a hydrographic survey obtained in January 2017, at three hydrographic stations along the Ganpu Transect, measuring tidal levels, water depth and current velocity for different water depths. Covering a spring-neap tide cycle, the measurements show that the tidal is variable in time and space \cite{huang2022}.

Tidal bores were generated using two methods:
\begin{enumerate}
\item \textbf{Partially closing vertical gate} (as in \cite{thomas2022}): a motor-driven gate falls with controlled speed, abruptly reducing the cross-section and generating a bore.
\item \textbf{Pump-stopping method} (as in \cite{rousseaux2016}): the pump is halted, allowing the downstream reservoir to fill until hydraulic inversion generates a surge.
\end{enumerate}

The velocity fields are measured using PIV method, which provides high‑resolution, two component velocity maps in a vertical plane. A laser sheet illuminates neutrally buoyant seeding particles, and a high speed camera records their motion. The PIV processing yields the horizontal and vertical velocity components, as well as streamlines that reveal the flow structures beneath the bore.

Although PIV is not designed to measure the free surface directly, the surface location can be inferred from the absence of particles and from the deformation of the illuminated region. This allows a qualitative comparison between the experimental and numerical free surface profiles (more details are presented in the appendix and Fig. \ref{TBPIV}).

\begin{figure}
\centering
\includegraphics[width=0.4\textwidth]{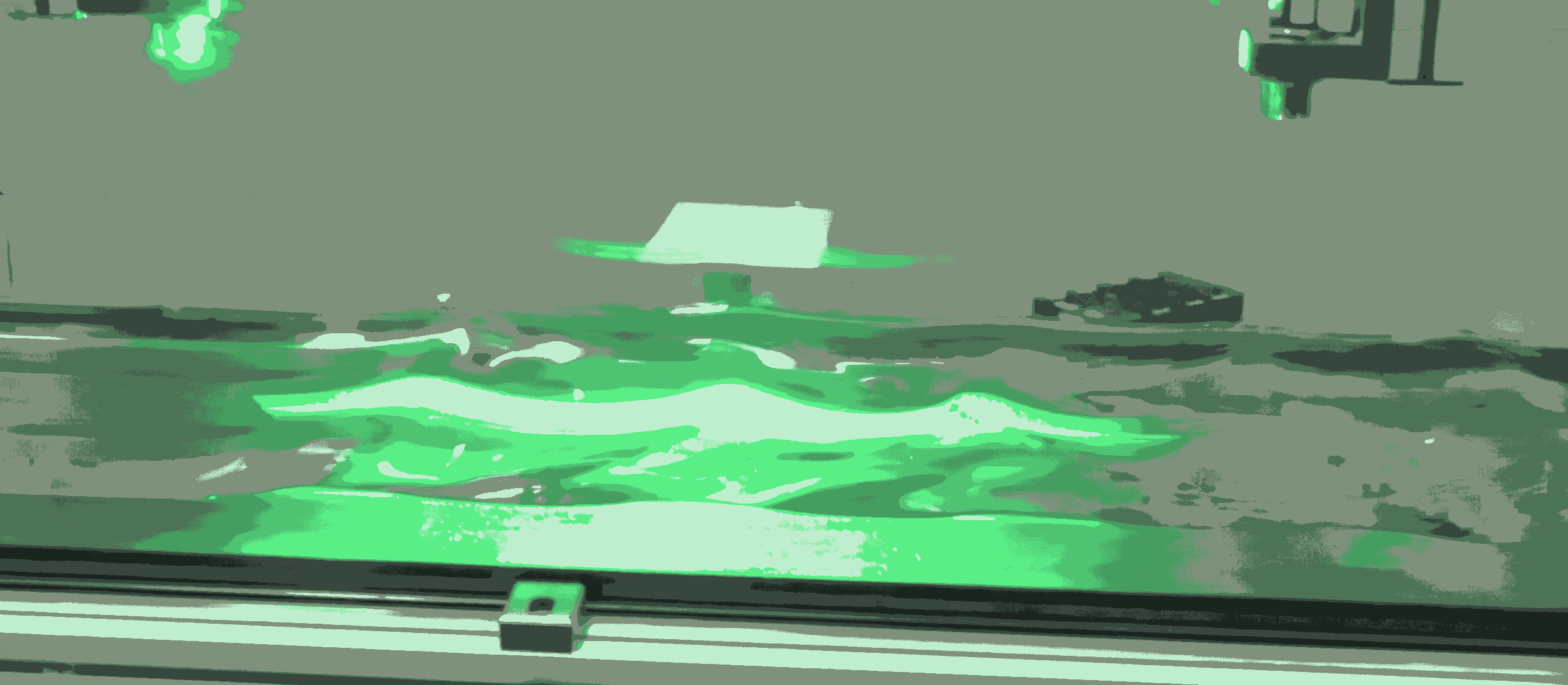}
\caption{Perspective view of the light‑green LASER sheet illuminating the water column and delineating the bore front together with its first two secondary undulations (“whelps”). Two Microsonic ultrasonic sensors, used to monitor the mean water level, are visible at the top of the frame. The background current flows from left to right.}
\label{TBPIV}
\end{figure}

\section{Main results}

\subsection{Presentation of the study cases}

The system (bore + river) is described by five physical variables: $h_1$ and $V_1$ for the depth and velocity respectively upstream, $h_2$ and $V_2$ for the depth and velocity respectively downstream, and $V_b$ the bore velocity. In this study, the horizontal component of the bore velocity $V_b$ is estimated by following the bore front in both the numerical simulations and laboratory experiments. The estimated value is quite constant for the whelps during the bore passing and is always opposite to $V_1$. As for the velocity after the bore passing $V_2$, it can initially be either positive or negative with respect to $V_1$. $h_2$ is always higher than $h_1$ in order to have a positive surge. 
The two non-dimensional numbers: Froude number ($Fr$) and Reynolds number ($Re$), are used to characterize the undular bores.

The tidal range is inhomogenous in the experiments, the resulting water depth after the bore passing $h_2$ varies in time, which is not the case in the numerical simulation where $h_2$ is constant. This leads to evolving tidal bores during the experiments \cite{rousseaux2016}. This will explain the classification considered for the first case initially representing a case B which turned out into a case A. As we want to have one examplary of each case reported in this work, we decided arbitrarily to classify the latter example as a case A. Similarly, the third case initially gives a case C which turned out into a case D. Again, we decided to report this other example as a case C in our study. These observations show that the downstream depth in the channel or the post-bore depth in the estuary impacts the classification between the different cases A, B, C, and D: one must be very careful as one can have transition between cases for an evolving tidal range in a given experiment contrary to the numerical simulations where the gate forces a constant tidal range as we will show later in the paper.}

\begin{table*}[ht]
\caption{Initialisation parameters of 2D numerical simulations for the four cases.}
\centering
\begin{tabular}{|c|c|c|c|c|c|c|}
\hline
Case & $L$ $(m)$ & $X_g$ $(m)$ & $Z_g$ $(m)$ & $V_g$ $(m.s^{-1})$ & $t_g$ $(s)$ & $h_g$ $(m)$ \\
\hline
A & $17$ & $11$ & $0.0015$ & $0.5$ & $0.27$ & $0.014$ \\
B & $17$ & $11$ & $0.0015$ & $0.5$ & $0.25$ & $0.023$ \\
C & $17$ & $11$ & $0.0015$ & $0.5$ & $0.3$ & $0$ \\
D & $17$ & $11$ & $0.0024$ & $0.5$ & $0.48$ & $0$ \\
\hline
\end{tabular}
\label{table1}
\end{table*}

\begin{table*}[ht]
\caption{Physical parameters of 2D numerical simulations for the four cases.}
\centering
\begin{tabular}{|c|c|c|c|c|c|c|c|c|}
\hline
Case & $V_1$ $(m.s^{-1})$ & $h_1$ $(m)$ & $V_2$ $(m.s^{-1})$ & $h_2$ $(m)$ & $V_b$ $(m.s^{-1})$ & $V_b / V_1$ & $Fr$ & $Re$ \\
\hline
A & $0.32$ & $0.097$ & $0.09$ & $0.128$ & $0.89$ & $2.78$ & $1.24$ & $31,040$ \\
B & $0.41$ & $0.098$ & $0.21$ & $0.13$ & $0.8$ & $1.95$ &$1.23$ & $39,360$ \\
C & $0.28$ & $0.096$ & $0$ & $0.135$ & $0.95$ & $3.39$ & $1.27$ & $26,880$ \\
D & $0.28$ & $0.15$ & $0$ & $0.2$ & $1.2$ & $4.29$ &$1.22$ & $42,000$ \\
\hline
\end{tabular}
\label{table2}
\end{table*}

Using the previous numerical model with the closing gate method only (see Fig. \ref{fig:geometry}), the four cases of the classification of undular bores introduced by Berchet $et \, al.$ in 2018 \cite{berchet2018flow} were simulated (see Fig. \ref{TBclass2}). In the numerical simulation, the flow was initialised in an open channel of length $L$ with an uniform velocity profile $V_1$ and a water depth $h_1$. The flow direction was from left to right. The gate was located at the position $(X_g, Z_g)$. The gate closure was defined by a falling speed $V_g$ for a duration $t_g$ and a remaining opening $h_g$. The initial and physical parameters of the numerical simulation are summarized in Tables \ref{table1} and \ref{table2} for the four studied cases. The four cases were also observed in the experiments albeit sometimes with a partial wave-breaking that does not impinged the conclusions reported in this work. The physical parameters used in the experiments are summarized in Table \ref{table3}. The flow direction, in the experiments, was also from left to right (see Fig. \ref{bore-setup}). 

For the four cases, the simulated real time was fixed to $7\,s$. For the three first cases, A, B and C, the particles number was the same, i.e $282,047$ particles. In the last case D, 450,075 particles were used. There were more particles in this case as the initial water elevation was increased. For the GPU time, the cases, A, B and C consumed on average $12\,min$. Using more particles, the case D has required more GPU time, $35\,min$. The numerical simulation, here presented, is performed with a graphical card NVIDIA Corporation GK 104 [GeForce GTX 660 Ti] (rev a1) [\url{https://www.nvidia.com/content/pdf/geforce-gtx-660ti.pdf}] with the language Cuda version 11.4 NVIDIA-SMI 470.256.02 [\url{https://docs.nvidia.com/datacenter/tesla/tesla-release-notes-470-256-02/index.html}].

\begin{table*}[ht]
\caption{Physical parameters of the experiments for the four cases.}
\centering
\begin{tabular}{|c|c|c|c|c|c|c|c|c|c|}
\hline
	Case & $V_1$ $(m.s^{-1})$ & $h_1$ $(m)$ & $V_2$ $(m.s^{-1})$ & $h_2$ $(m)$ & $V_b$ $(m.s^{-1})$ & $a_d$ $(m.s^{-2})$ & $Fr$ & $Re$ & Technic used \\
\hline
	A & $0.61$ & $0.094$ & $--$ & $--$ & $0.59$ & $-0.02$ & $1.25$ & $57,340$ & Partially closed gate \\
	B & $0.6$ & $0.094$ & $--$ & $--$ & $0.72$ & $-0.018$ & $1.37$ & $56,400$ & Partially closed gate \\
	C & $0.34$ & $0.047$ & $--$ & $--$ & $0.63$ & $-0.018$ & $1.43$ & $15,980$ & Pump stopping \\
	D & $0.28$ & $0.043$ & $--$ & $--$ & $0.75$ & $-0.018$ & $1.58$ & $12,040$ & Pump stopping \\
\hline
\end{tabular}
\label{table3}
\end{table*}

The identification of a flow reversal is then performed by analysing the flow velocity $\bf V$ (black vector) under the bore front with an observation window (see Fig. \ref{fig:figuresoil1} for the case A, Fig. \ref{fig:figuresoil2} for the case B, Fig. \ref{fig:figuresoil3} for the case C, Fig. \ref{fig:figuresoil4} for the case D, top: numerical simulation and bottom: experiment). The size of the observation window was fixed between $x = -0.1\,m$ and $x = 0.45\,m$ for the numerical simulation and between $x = -0.24\,m$ and $x = 0.26\,m$ for the experiment. An analysis of the non-dimensional value of the horizontal velocity component $V_x/V_1$ at an observation point with respect to a non-dimensional time $t\sqrt{g/h_1}$ (see Fig. \ref{fig:profilA} for the case A, Fig. \ref{fig:profilB} for the case B, Fig. \ref{fig:profilC} for the case C, Fig. \ref{fig:profilD} for the case D, top: experiment, and bottom: numerical simulation) is also done. The observation point was fixed at the position ($X=4\, m$, $Z=0.06\, m$) for the numerical simulation and at the positions: ($X=-0.23\, m$, $Z=0.06\,m$) in cases A and B, ($X=-0.18\, m$, $Z=0.022\,m$) in case C, and ($X=0.032\, m$, $Z=0.029\,m$) in case D, for the experiment in the water column. Considering the difference between the experiment and simulated conditions, the numerical simulation and the experiment show similar trend and evolution during the bore passing. The vertical component $V_z$ of the flow velocity is not presented here since it globally oscillates proportionally to the free surface evolution in both the cases studied. The streamlines of the wave front are given for all the cases in the three reference frames, except for case A in the experiment where the time $t = 10.936$ $s$ has been selected.

\subsection{Case A: hydrodynamics dominated by undulations}
\begin{figure}
\includegraphics[width=0.9\columnwidth]{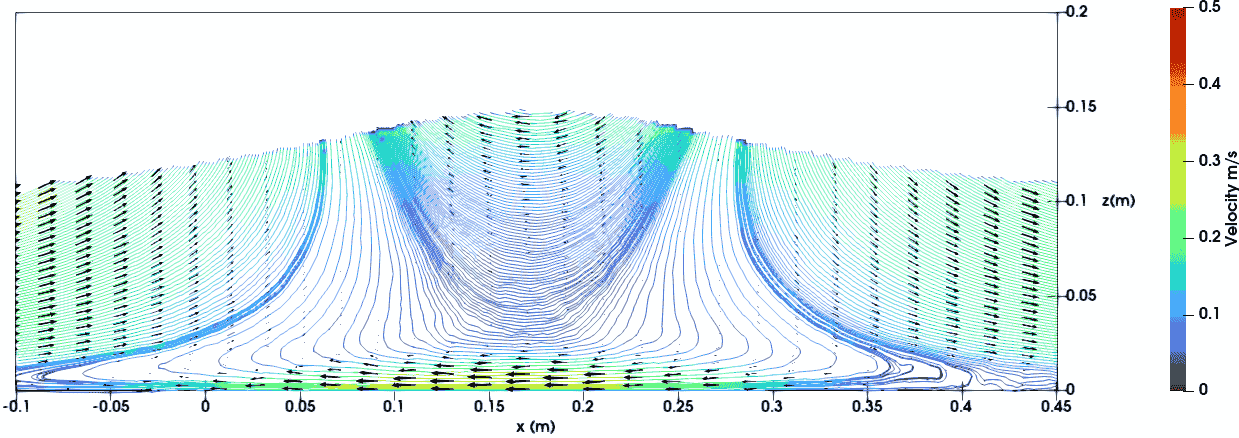}
\includegraphics[width=0.9\columnwidth]{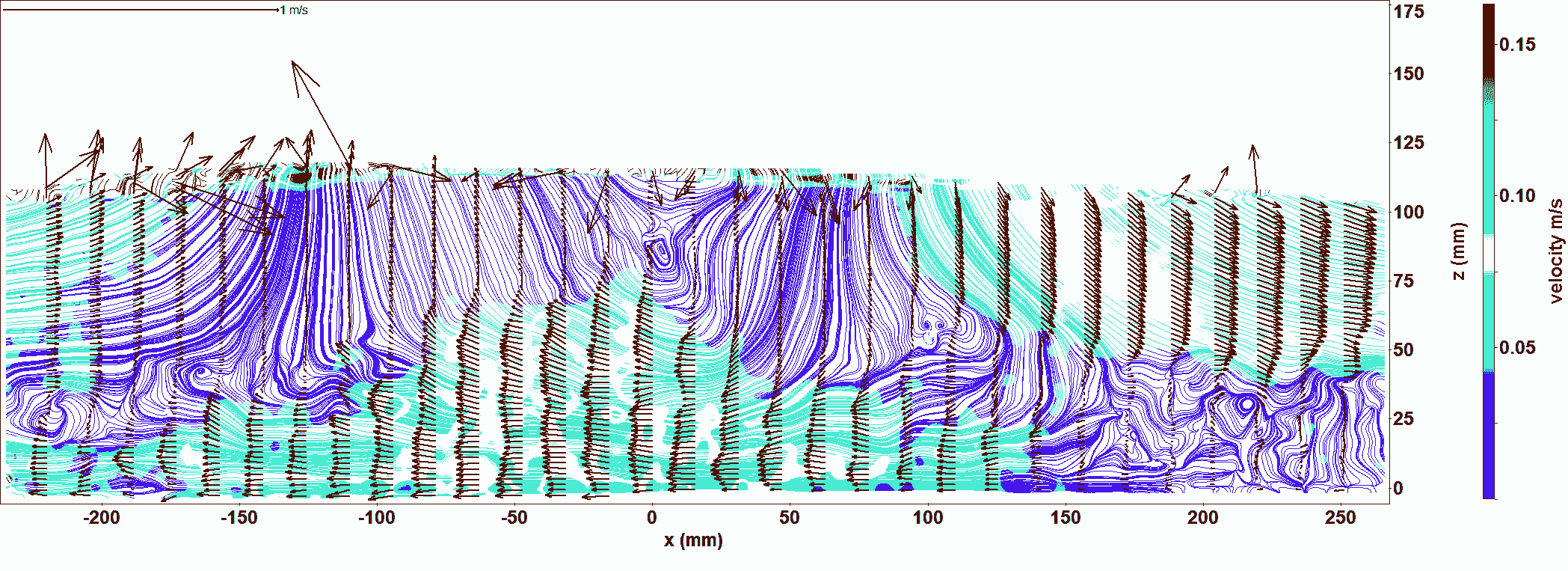}
	\caption{Case A observed in the numerical simulation (num) with $V_1=0.32\, m.s^{-1}$, $h_1=0.097\,m$, $Fr = 1.24$ and $Re = 31, 040$ using Paraview at $t=7\,s$ (top), and in the experiment (exp) using the partially closed gate method with $V_1=0.61\, m.s^{-1}$, $h_1=0.094\,m$, $Fr = 1.25$ and $Re=57, 340$ using DaVis8 at $t=10.936\,s$, $im=2734$ (bottom). In both figures, the bore is undular (num) / undular breaking (exp) and exhibits a partial flow reversal beneath the first crest, particularly near the bed. The streamlines reveal an "elephant‑foot" recirculation structure characteristic of Cases A or C dynamics. No global flow reversal is observed after the passage of many whelps but only with the decaying tidal range ultimately.}
\label{fig:figuresoil1}
\end{figure}

In this case, the flow direction alternates under the bore front (see Fig. \ref{fig:figuresoil1}). The flow velocity $\bf V$ is in the downstream direction before and after the wave crest, and it is in the upstream direction in the wave crest. The inversion of the flow direction is partial along the water column under the wave crest in both the numerical simulation (see Fig. \ref{fig:figuresoil1} - top) and the experiment (see Fig. \ref{fig:figuresoil1} - bottom; the image of the taken snapshot is also given). The numerical simulation shows the bore front as for the experiment a snapshot of a whelp is shown. Under the bore front and close to the bed, the flow reversal was particularly strong near the bed. The streamlines show an elephant foot under the crest, this, in both the numerical simulation (see Fig. \ref{fig:figuresoil1} - top) and the experiment (see Fig. \ref{fig:figuresoil1} - bottom). The Froude number of the numerical simulation was very similar to the one of the experiment, $Fr$ = 1.24 and 1.25 for the numerical simulation and the experiment, respectively (see Tab. \ref{table2} and \ref{table3}). The tidal bore is undular.

\begin{figure}
\includegraphics[width=0.45\textwidth]{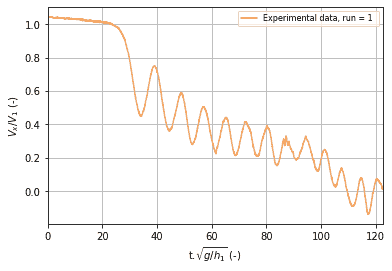}
\includegraphics[width=0.45\textwidth]{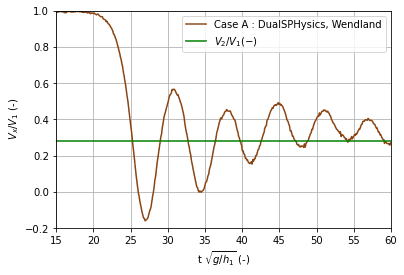}
	\caption{Non-dimensional value of the horizontal velocity component ${V_x}/{V_1}$ as a function of the non-dimensional time $t\sqrt{g/{h_1}}$ of case A at the position ($X=4\,m$, $Z=0.06\,m$) for the numerical simulation (right) and at the position ($X=-0.23\,m$, $Z=0.06\,m$) for the experiment using the partially closed-gate method (left). In both datasets, the velocity oscillates in phase with the free surface undulations, displaying alternating acceleration and deceleration with progressively decreasing amplitude. A partial flow reversal occurs beneath the wave front in the simulation, while in the experiment the velocity becomes weakly negative only near the end of the time series due to the gradual drop in water level.}
\label{fig:profilA}
\end{figure}

In Fig. \ref{fig:profilA}, the non-dimensional value of the horizontal velocity component ${V_x}/{V_1}$ closely follows the free surface evolution during the bore passing in both the numerical simulation and the experiment. As the water level oscillates, it alternatively decelerates and accelerates with a velocity range progressively decreasing. The horizontal component $V_x$ of the flow velocity oscillates around the flow velocity $V_2$ after the bore passing, $V_2/V_1$ = 0.28 in the numerical simulation (see Fig. \ref{fig:profilA} - bottom). In the numerical simulation, $V_x$ changes direction flowing upstream, negative value, beneath the wave front. Beneath the frontal wave trough, the flow direction changes again, flowing downstream. Beneath the whelps, the flow direction does not change, no negative values of $V_x$. In the experiment, the non-dimensional value of the horizontal velocity component ${V_x}/{V_1}$ is positive except at the end (region between 100 $mm$ and 120 $mm$) where it alternate between a positive and a negative value. The flow reversal is partial in both the numerical simulation and the experiment.
In case A, the bore velocity is numericaly estimated to $0.89$ $m.s^{-1}$. The ratio $V_b / V_1$ is then equal to $2.78$ with the upstream velocity $V_1$ = $0.32$ $m.s^{-1}$.

\subsection{Case B: hydrodynamics dominated by the initial current}
\begin{figure}
\includegraphics[width=0.9\columnwidth]{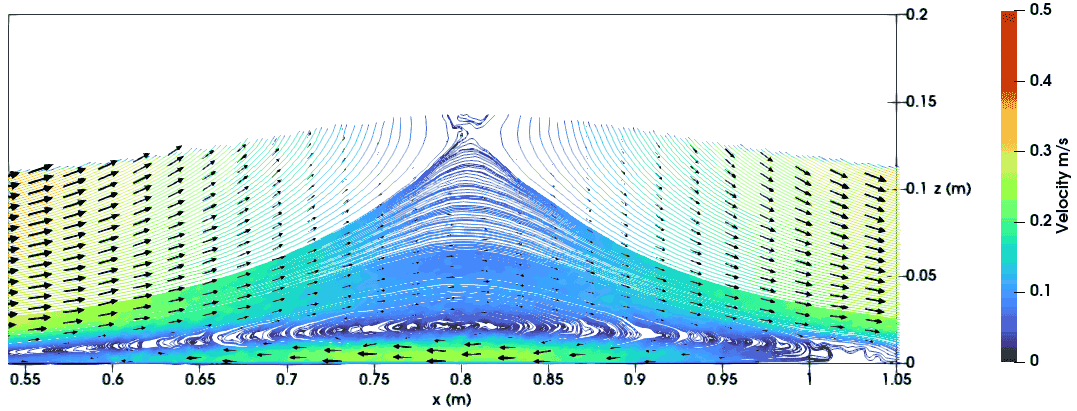}
\includegraphics[width=0.9\columnwidth]{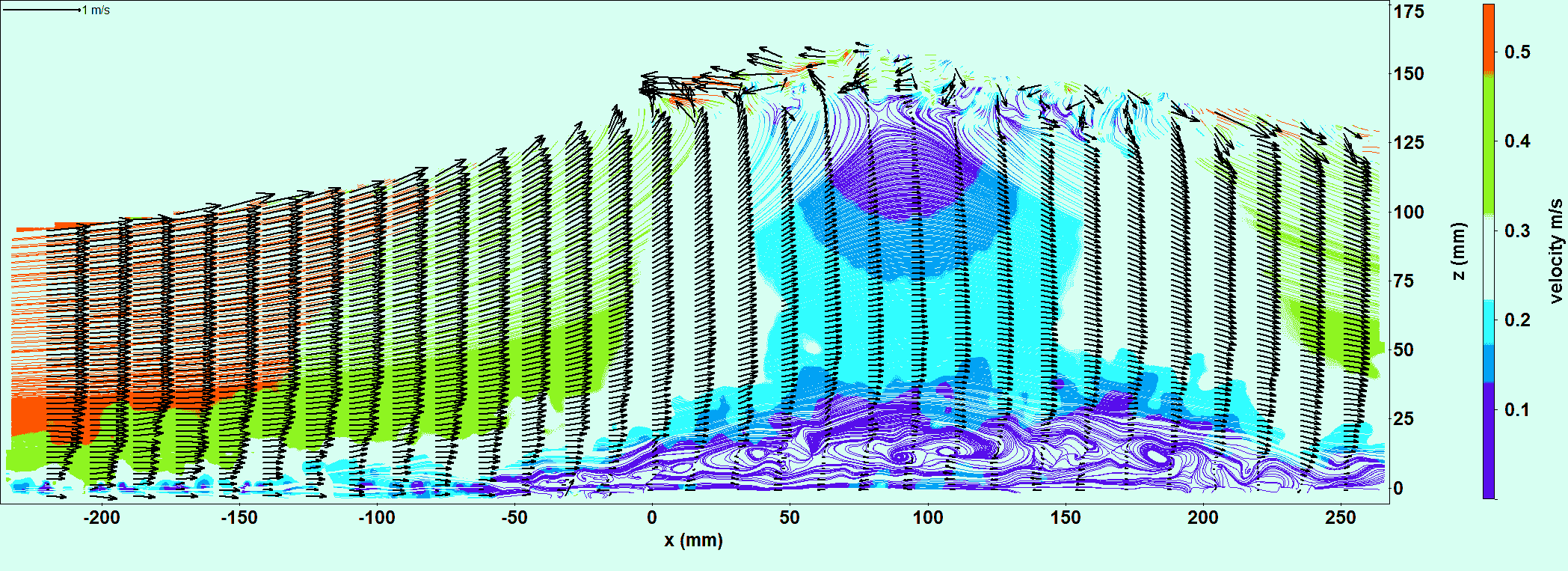} 
	\caption{Case B observed in the numerical simulation (num) with $V_1=0.41\, m.s^{-1}$, $h_1=0.096\,m$, $Fr = 1.24 $ and $Re = 39, 360$ using Paraview at $t=7\,s$ (top), and the experiment (exp) using the partially closed gate method with $V_1=0.6\, m.s^{-1}$, $h_1=0.094\,m$, $Fr = 1.37$ and $Re = 56,400$ using DaVis8 at $t=3.292\,s$, $im=823$ (bottom). In the figures, the bore is undular (num) / undular breaking (exp) and induces a marked deceleration beneath the front without causing a reversal of the main flow direction, except for a thin near‑bed layer. Streamlines reveal a recirculation bubble under the crest, consistent with Case B dynamics.}
\label{fig:figuresoil2}
\end{figure}

In Fig. \ref{fig:figuresoil2}, one can observe that the numerical simulation and the experiment gave the same hydrodynamics for the water flow despite a difference of the Froude number between the two, $Fr$ = 1.24 and 1.37 for the numerical simulation and the experiment, respectively (see Tab. \ref{table2} and \ref{table3}). The numerical simulation (see Fig. \ref{fig:figuresoil2} - top) and the experiment (see Fig. \ref{fig:figuresoil2} - bottom) give an undular tidal bore. One has high values of the flow velocity $\bf V$ just before the bore passage. $\bf V$ is zero near the bed area and a small area near the water column of the wave front. Under the frontal wave, the bore produces a flow deceleration with no change of direction in both the numerical simulation (see Fig. \ref{fig:figuresoil2} - top) and the experiment (see Fig. \ref{fig:figuresoil2} - bottom). The flow velocity $\bf V$ was in the downstream direction except near the bed area and near the water column. Under the wave front, beneath the bed, an intense flow reversal appears. From the streamlines, one can observe the recirculation bubble under the crest in the numerical simulation (see Fig. \ref{fig:figuresoil2} - top) and the experiment (see Fig. \ref{fig:figuresoil2} - bottom). 

\begin{figure}
\includegraphics[width=0.45\textwidth]{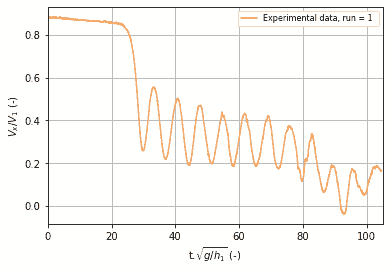}
\includegraphics[width=0.45\textwidth]{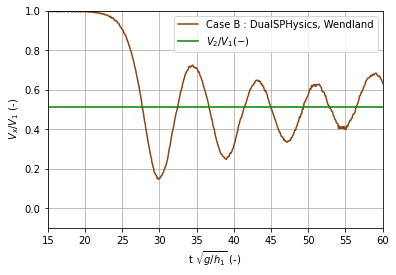}
	\caption{Non-dimensional value of the horizontal component ${V_x}/{V_1}$ of the flow velocity as a function of non-dimensional time $t\sqrt{g/{h_1}}$ of case B at the position ($X=4\,m$, $Z=0.06\,m$) for the numerical simulation (right) and at the position ($X=-0.23\,m$, $Z=0.06\,m$) for the experiment using the partially closed gate method (left). In both datasets, the velocity oscillates in phase with the free surface undulations, showing alternating deceleration beneath the crest and acceleration beneath the trough, without sign reversal—unlike Case A. The oscillation amplitude gradually decreases downstream. The slight decline in measured velocities in the experiment is attributed to the progressive drop in water level during the run.}
\label{fig:profilB}
\end{figure}

In Fig. \ref{fig:profilB}, the non-dimensional value of the horizontal velocity component $V_x/V_1$ evolves in synchronization with the free surface in both the numerical simulation and the experiment. As the water level oscillates, it alternatively decelerates and accelerates beneath the wave crest and wave trough, respectively contrarily to the case A where $V_x/V_1$ changes direction. The horizontal component $V_x$ of the flow velocity oscillates around the flow velocity $V_2$ after the bore passing, $V_2/V_1$ = 0.51 in the numerical simulation (see Fig. \ref{fig:profilB} - bottom). There is no complete change of direction of the current over the water column in both the numerical simulation (see Fig. \ref{fig:profilB} - bottom) and the experiment (see Fig. \ref{fig:profilB} - top). $V_x$ remains in the same direction as the flow velocity $V_1$ of the river before the bore passing in most of the water column. The velocity range progressively decreases. At the difference of the case A, there is no flow reversal. $V_x$ of the flow velocity does not change its sign. As in the case A, a decrease of the horizontal velocity component is observed in the experiment. This is due to the inhomogeneity of the water depth $h_2$ after the bore passing in this experiment. 
In case B, the bore velocity is numericaly estimated to $0.8$ $m.s^{-1}$. The ratio $V_b / V_1$ is then equal to $1.95$ with the upstream velocity $V_1$ = $0.41$ $m.s^{-1}$.

\subsection{Case C: hydrodynamics partially dominated by undulations and then the bore flow}
\begin{figure}
\includegraphics[width=0.9\columnwidth]{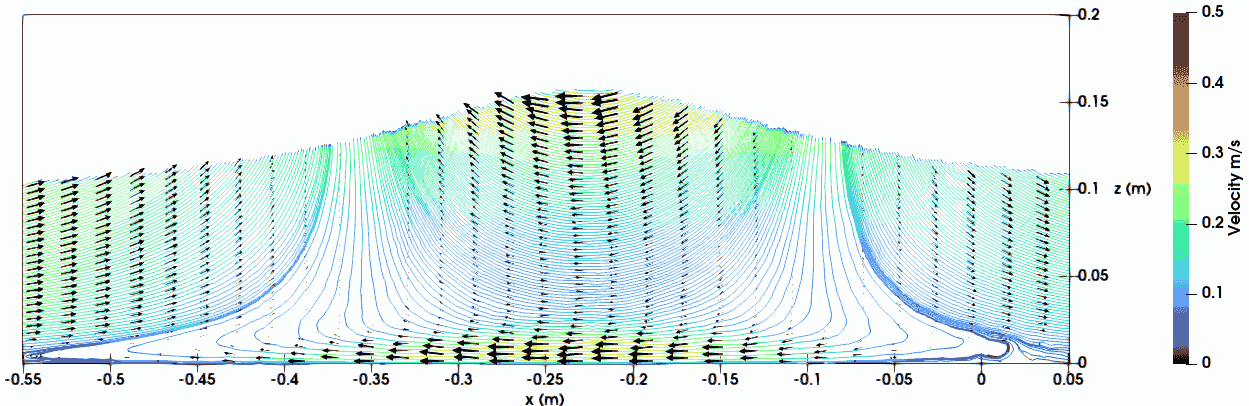}
\includegraphics[width=0.9\columnwidth]{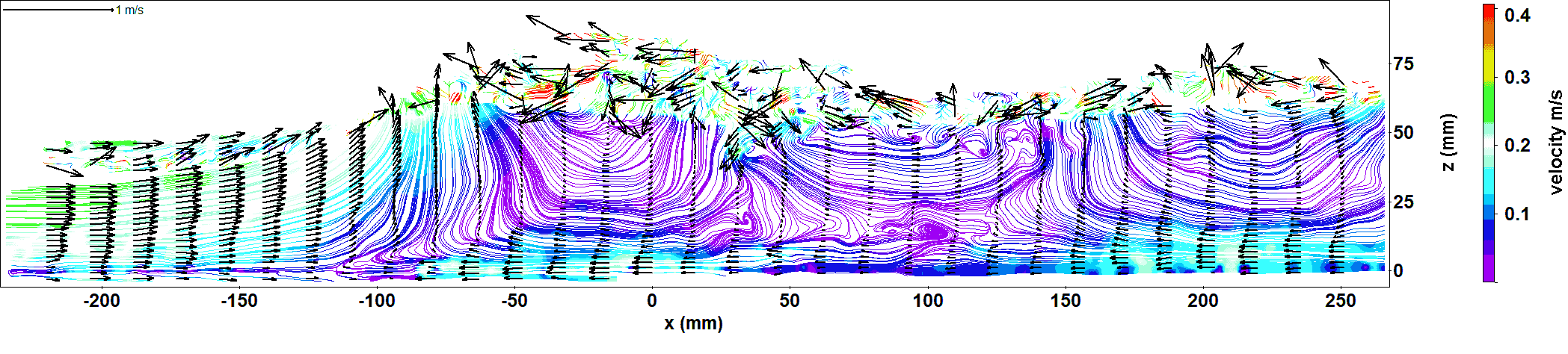}
	\caption{Case C observed in the numerical simulation (num) with $V_1=0.28\, m.s^{-1}$, $h_1=0.096\,m$, $Fr = 1.26$ and $Re = 26, 880$ using Paraview at $t=7\,s$ (top), and the experiment using the pump stopping method with $V_1=0.34\, m.s^{-1}$, $h_1=0.047\,m$, $Fr = 1.43$ and $Re = 15, 980$ using DaVis8 at $t=9.024\,s$, $im=2256$ (bottom). Both simulation and experiment exhibit an undular (num) / undular breaking (exp ) bore with a clear flow reversal beneath the wave front, extending across the full water column. The streamlines reveal an "elephant‑foot" recirculation structure characteristic of Cases A or C dynamics. A global flow reversal is observed.}
\label{fig:figuresoil3}
\end{figure}

In this case, the Froude number $Fr$ is equal to 1.26 and 1.43 for the numerical simulation and the experiment, respectively (see Tab. \ref{table2} and \ref{table3}). In Fig. \ref{fig:figuresoil3}, the bore is undular, and this, in both the numerical simulation and the experiment. One can observe that the flow velocity $\bf V$ alternates its direction under the crest in all the water column, in both the numerical simulation (see Fig. \ref{fig:figuresoil3} - top) and the experiment (see Fig. \ref{fig:figuresoil3} - bottom). The flow velocity $\bf V$ is in the downstream direction before and after the crest, and it is in the upstream direction in the crest. It indicates the presence of a flow reversal. The streamlines show also an elephant foot under the wave front in both the numerical simulation (see Fig. \ref{fig:figuresoil3} - top) and the experiment (see Fig. \ref{fig:figuresoil3} - bottom). Unlike the case A, the flow velocity $\bf V$ is higher in the crest, higher near the bed area and higher near the water column.

\begin{figure}
\includegraphics[width=0.45\textwidth]{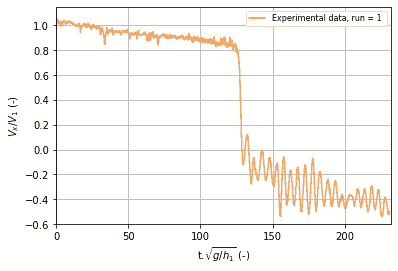}
\includegraphics[width=0.45\textwidth]{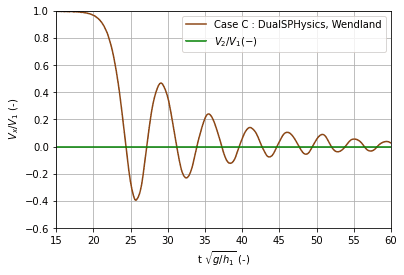}
	\caption{Non-dimensional value of the horizontal component ${V_x}/{V_1}$ of the flow velocity as a function of the non-dimensional time $t\sqrt{g/{h_1}}$ of the case C at the position ($X=4\,m$, $Z=0.06\,m$) for the numerical simulation (right) and at the position ($X=-0.18\,m$, $Z=0.022\,m$) for the experiment using the pump stopping method (left). In both datasets, the velocity oscillates in phase with the free surface undulations, showing alternating acceleration and deceleration with decreasing amplitude downstream. A partial flow reversal occurs beneath the bore front and persists over several subsequent whelps. In the experiment, the negative values of $V_x/V_1$ between 125 $mm$ and 225 $mm$ confirm the transition toward Case D–like behaviour. The overall decline in measured velocities reflects the gradual drop in water level during the run.}
\label{fig:profilC}
\end{figure}

In Fig. \ref{fig:profilC}, the non-dimensional value of the horizontal component $V_x/V_1$ of the flow velocity closely follows the free surface evolution during the bore passing in both the numerical simulation and the experiment. As the water level oscillates, it alternatively decelerates and accelerates with a velocity range progressively decreasing. The horizontal component $V_x$ of the flow velocity oscillates around the flow velocity $V_2$ after the bore passing, $V_2/V_1$ = 0 in the numerical simulation (see Fig. \ref{fig:profilC} - bottom). In the numerical simulation, $V_x$ changes direction flowing upstream, negative value, beneath the frontal wave. Beneath the  wave front trough, the flow direction changes again, flowing downstream. Beneath the whelps, a similar pattern took place. In the experiment, the non-dimensional value of the horizontal velocity component ${V_x}/{V_1}$ becomes negative in the region between 125 $mm$ and 225 $mm$ (see Fig. \ref{fig:profilC} - top). This shows that the hydrodynamics of the flow changes from case A to case D. The observed decrease of the horizontal velocity component is due to a dropping of the water level in the experiment. The flow reversal is partial in both the numerical simulation (see Fig. \ref{fig:profilC} - bottom) and the experiment (see Fig. \ref{fig:profilC} - top), not total.  
In case C, the bore velocity is numericaly estimated to $0.95$ $m.s^{-1}$. The ratio $V_b / V_1$ is then equal to $3.39$ with the upstream velocity $V_1$ = $0.28$ $m.s^{-1}$.

\subsection{Case D: hydrodynamics dominated by the bore flow}
\begin{figure}
\includegraphics[width=0.9\columnwidth]{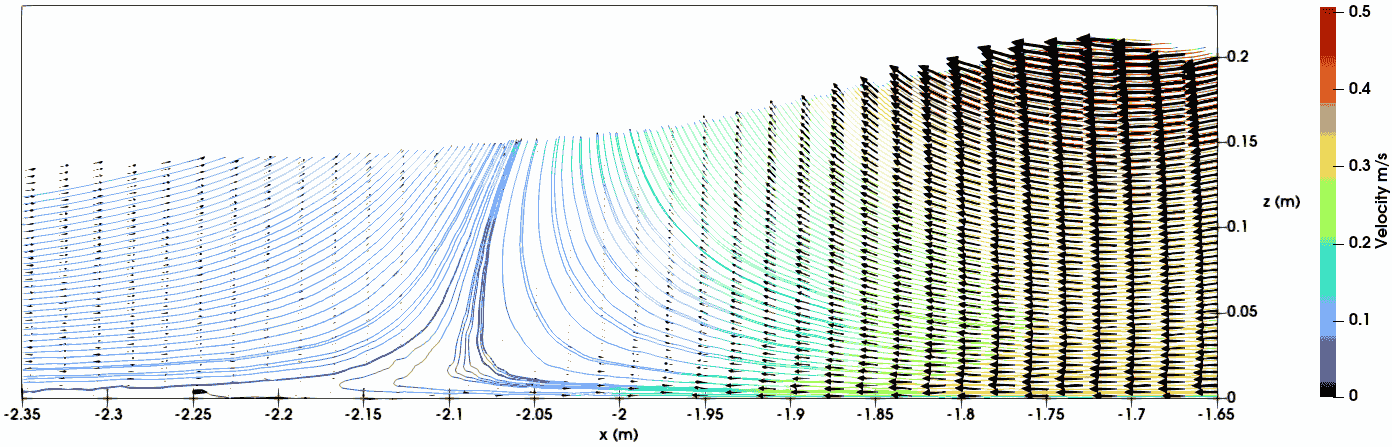}
\includegraphics[width=0.9\columnwidth]{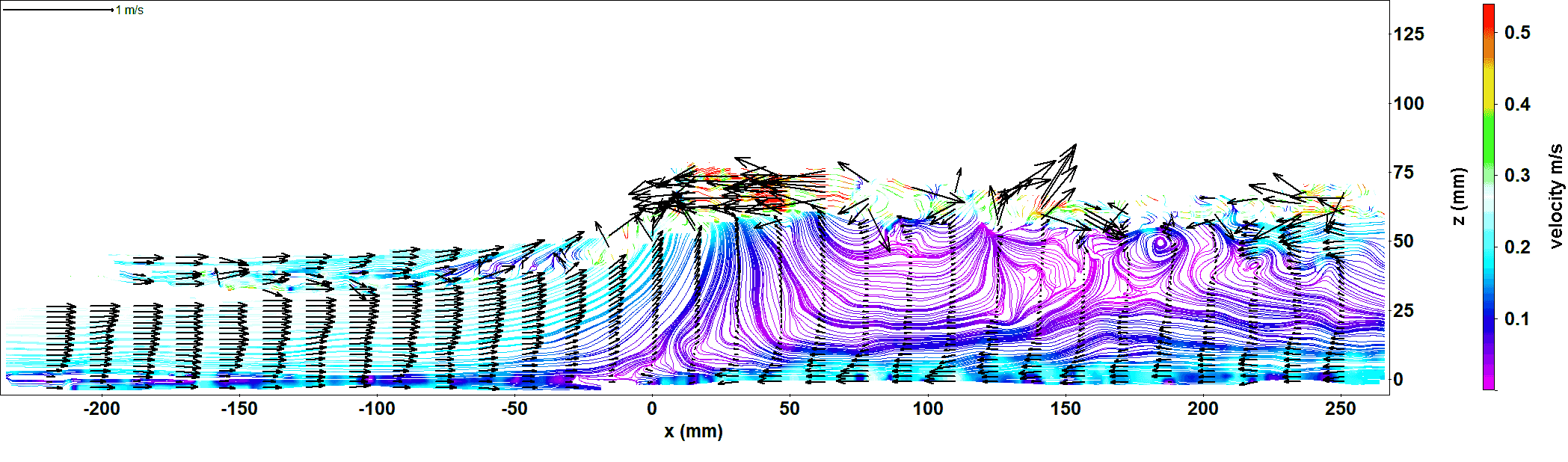}
	\caption{Case D observed in the numerical simulation obtained with $V_1=0.28\, m.s^{-1}$, $h_1=0.15\,m$, $Fr = 1.22$ and $Re=42, 000$ using Paraview at $t=7\,s$ (top) and the experiment using the pump stopping method with $V_1=0.28\, m.s^{-1}$, $h_1=0.043\,m$, $Fr = 1.58$ and $Re=12, 040$ using DaVis8 at $t=13.968\,s$, $im=3942$ (bottom). The numerical bore remains undular, whereas the experimental bore exhibits a partially breaking crest. In both cases, the flow undergoes a strong and nearly column‑wide reversal beneath the bore front. The velocity magnitude is highest under the wave front, with intense upstream‑directed motion near both the free surface and the bed. Streamlines reveal a pronounced, fully developed vertical boundary characteristic of Case D reversal dynamics at the front passage.}
\label{fig:figuresoil4}
\end{figure}

In this case, the Froude number $Fr$ is equal to 1.22 and 1.58 for the numerical simulation and the experiment, respectively (see Tab. \ref{table2} and \ref{table3}). In Fig. \ref{fig:figuresoil4}, the bore is undular in the numerical simulation and partially breaking, a breaking crest (see Fig. \ref{fig:figuresoil4} - bottom), in the experiment. One can observe that the flow velocity $\bf V$ changes its direction under the crest in all the water column, in both the numerical simulation (see Fig. \ref{fig:figuresoil4} - top) and the experiment (see Fig. \ref{fig:figuresoil4} - bottom). After the crest, the flow velocity $\bf V$ is still in the upstream direction except in some areas near the water column. This indicates the presence of a flow reversal in both the numerical simulation and the experiment. In Fig. \ref{fig:figuresoil4}, one can observe that the flow velocity $\bf V$ is larger beneath the wave front than after during the bore passing. It is high in all the water column under the wave front. Under the frontal wave, the velocity is intense both near the free surface and near the bed than in the rest of the water column.  The streamlines show a clear flow reversal beneath the frontal wave. The comparison between the numerical simulations and the experiments shows that the reversal classification is restricted to undular and weakly breaking bores. 

\begin{figure}[htbp]
\includegraphics[width=0.45\textwidth]{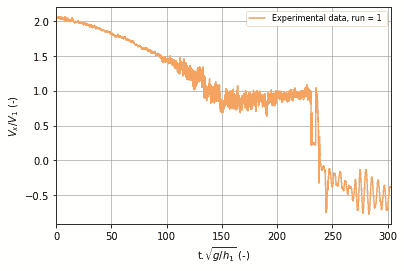}
\includegraphics[width=0.45\textwidth]{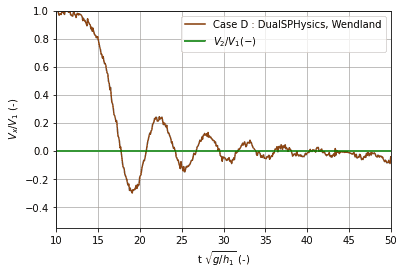}
	\caption{Non-dimensional value of the horizontal component ${V_x}/{V_1}$ of the flow velocity as a function of the non-dimensional time $t\sqrt{g/{h_1}}$ of the case D at the position ($X=4\,m$, $Z=0.06\,m$) for the numerical simulation (right) and at the position ($X=0.032\,m$, $Z=0.029\,m$) for the experiment using the pump stopping method (bottom). Both datasets show a sharp and complete flow reversal at the bore front. $V_x/V_1$ switches from positive to strongly negative values across the full water column. Subsequent undulations produce velocity oscillations that remain entirely in the negative region, indicating a sustained upstream‑directed flow. The decrease in the experimental velocity amplitude is attributed to the gradual lowering of the water level during the run.}
\label{fig:profilD}
\end{figure}

For this case, the flow suddenly changed direction beneath the wave front and undulated regularly after (see Fig. \ref{fig:profilD}). The non-dimensional value of the horizontal component $V_x/V_1$ of the flow velocity closely follows the free surface evolution during the bore passing in both the numerical simulation and the experiment. As the water level oscillates, it alternatively decelerates and accelerates with a progressively decreasing velocity range. In the experiment, even though the bore is partially breaking, the same observation is made (see Fig. \ref{fig:profilD} - top). There is a complete flow reversal beneath the bore when $V_x/V_1$ changes its sign over the water column in both the numerical simulation and the experiment. During the wave front passage, $V_x/V_1$ goes from 1 to a negative value in the numerical simulation (see Fig. \ref{fig:profilD} - bottom). For the following secondary undulations, the horizontal velocity component $V_x$ followed a similar pattern to the one observed beneath the frontal wave and trough with a stabilization of the flow velocity around $V_2$. $V_2$ is equal to 0 $m.s^{-1}$ in the numerical simulation. Beneath the whelps, the velocity magnitudes oscillate in most of the water column and remains negative. This exhibits a complete flow reversal. In the experiment, the non-dimensional value of the horizontal velocity component ${V_x}/{V_1}$ suddenly becomes negative right after the bore passing (see Fig. \ref{fig:profilD} - top). In contrast to the case C, there is no region with a positive velocity values after the bore passing. After the wave front, the non-dimensional value of the horizontal velocity component ${V_x}/{V_1}$ oscillates in the negative region. This shows that the flow reversal is total. The observed decrease of the horizontal velocity component is due to a dropping of the water level in this experiment.
In case D, the bore velocity is numericaly estimated to $1.2$ $m.s^{-1}$. The ratio $V_b / V_1$ is then equal to $4.29$ with the upstream velocity $V_1$ = $0.28$ $m.s^{-1}$.

\section{Tidal bore and reference frames}

As indicated by kinematics, the movement of an objet is relative in nature \cite{calderon2021} \cite{steeneken2025}. Before it can be described, it is necessary to specify in relation to what the movement of this objet is considered, i.e. the reference frame \cite{doyle2009}.

In this section, we analyse the velocity fields obtained from numerical simulations in three frames of reference to tidal bore dynamics. This multi-frame analysis reveals that the classical classification of tidal bores based solely on free surface shape or on velocity patterns in the bank frame is incomplete. Several regimes that appear distinct in one frame become equivalent in another frame, and conversely, regimes that appear similar in the bank frame may correspond to fundamentally different flow structures when viewed from the river or bore frame. The combination of PIV measurements and WCSPH simulations allows us to reconstruct fully the velocity field and to transform it into any frame by a simple vector addition.

The three reference frames, $R_b$ river bank 
, $R_c$ river current 
and $R_m$ bore front
, are considered as Galilean reference frames in the numerical simulations and non Galilean reference frames in the experiments. In the numerical simulations and the experiments, the movements of the river and the tidal bore are mainly unidirectional. In the numerical simulations, the dynamcis of the river and the tidal bore are without deceleration of the flow velocity. The three reference frames satisfy Newton’s first law in the numerical simulations. In the experiments, the progressive drop in water level induces a deceleration of the flow velocity. The estimation of this deceleration is $a_d \approx -0.02$ $m.s^{-2}$ for the four cases considered in the present paper (see Tab. III). The ratio between this deceleration and the value of gravity is about $0.2\%$. As the inertial effects are low, the three reference frames are quasi-Galilean for the experiments. And the laws of decomposition of velocities and acceleration are kinematic laws independent of Galilean character, of physical nature, of reference frames.


1) In the estuaries worldwide where tidal bores occur, the study of the impact of tidal bores on turbulence production, sediment resuspension, and mixing processes, is performed in the bank frame \cite{chanson2011}. For examples, one can mention the recent works on the tidal bores, by Pan et al. (2026) \cite{pan2026} in the Qiantang River Estuary and by Arildsen et al. (2025) \cite{arildsen2025} in the Severn Estuary. In the bank frame, the river current flows downstream with velocity $V_1>0$, while the bore front propagates upstream with velocity $V_b<0$. The horizontal velocity field is therefore the superposition of the river current and the bore-induced flow. This physical quantity is responsible for turbulence production, sediment resuspension, and mixing processes. Resulting of the superposition of the river current and the bore-induced flow, its analyze is difficult to understand the impact of tidal bores alone on these different mechanisms present in the real estuarines system.

2) Transforming the velocity field into the river current frame consists of subtracting the uniform background velocity $V_1$. This frame is physically relevant for organisms, sediment particles, or tracers advected by the river flow.
In this frame:\\
- A global reversal corresponds to a negative velocity throughout the water column.\\
- A local reversal corresponds to negative velocities confined to the bore front or to the troughs of the first undulations.\\
- A bore that does not reverse the flow in the bank frame may still reverse it in the river frame if the bore induced velocity exceeds the river current locally.\\
This reveals a key result. Several bores classified as non-reversing in the bank frame actually produce local or global reversals in the river frame. This explains long-standing field observations where floats or surface sediment tracers exhibit downstream motion even when the bore propagates upstream (case B, see Fig. \ref{fig:figuresoil2}). It explains also why floats closed to the bed sometimes drift downstream while the floats on the free surface are still moving upstream as discussed by Partiot a long time ago \cite{partiot58, partiot61}.

The river current frame allows to simplify the study of the impact of undular or weakly breaking bores on the trajectories of organisms, sediment particles, or tracers adevected by the river flow. In the works by Berchet et al. (2018) \cite{berchet2018flow}, the trajectories of suspended particles, induced by undular bores with Froude numbers between 1.1 and 1.2, are numerically studied. The Chen’s model of wave-current interactions was used to study the trajectories of suspended particles simulated from the Maxey-Riley equation \cite{chen1964}. This allowed to identify the different components of these trajectoires: gravity, elevation and current slowing down at the bore passage, oscillations induced by the wave-current interaction. The calibration of Chen’s model required a tedious work.

3) The bore front frame is obtained by subtracting the bore velocity $V_b$. In this frame, the bore front is stationary and the river flows toward it. This frame reveals the intrinsic structure of the bore independently of its advection by the river current.
Several important features become clearer:\\
- The bore front always appears as an hydraulic jump in this frame.\\
- Undulations correspond to dispersive waves generated downstream of the jump.\\
- The transition between undular, breaking, and undular breaking regimes is more sharply defined.\\
The bore front frame can be useful for studying the features of tidal bores in the estuaries. In the Kampar Estuary, Wisha et al. (2022) determined the real-time properties of a tidal bore generation, hydraulic jump, and transport mechanism \cite{wisha2022}.  

Between the three reference frames, the law of decomposition of velocities is kinematic law independent of Galilean character, of physical nature, of reference frames \cite{calderon2021, steeneken2025}:
\begin{equation}
V_{x_{|_{R_m}}} = V_{x_{|_{R_b}}}  - (-V_{b})
\mbox{ and }
V_{x_{|_{R_c}}} = V_{x_{|_{R_b}}} - V_{1}
\end{equation}
with $V_{x_{|_{R_m}}}$ the water speed in the reference frame of the bore front $R_m$, $V_{x_{|_{R_c}}}$ the water speed in the reference frame of the river current $R_c$ and $V_{x_{|_{R_b}}}$ the water speed in the reference frame of the river banks $R_b$. In the first expression, the minus sign of the bore velocity $V_b$ is induced by the opposite movement between the river current and the tidal bore.

\begin{figure*}
\centering
\includegraphics[scale=0.1]{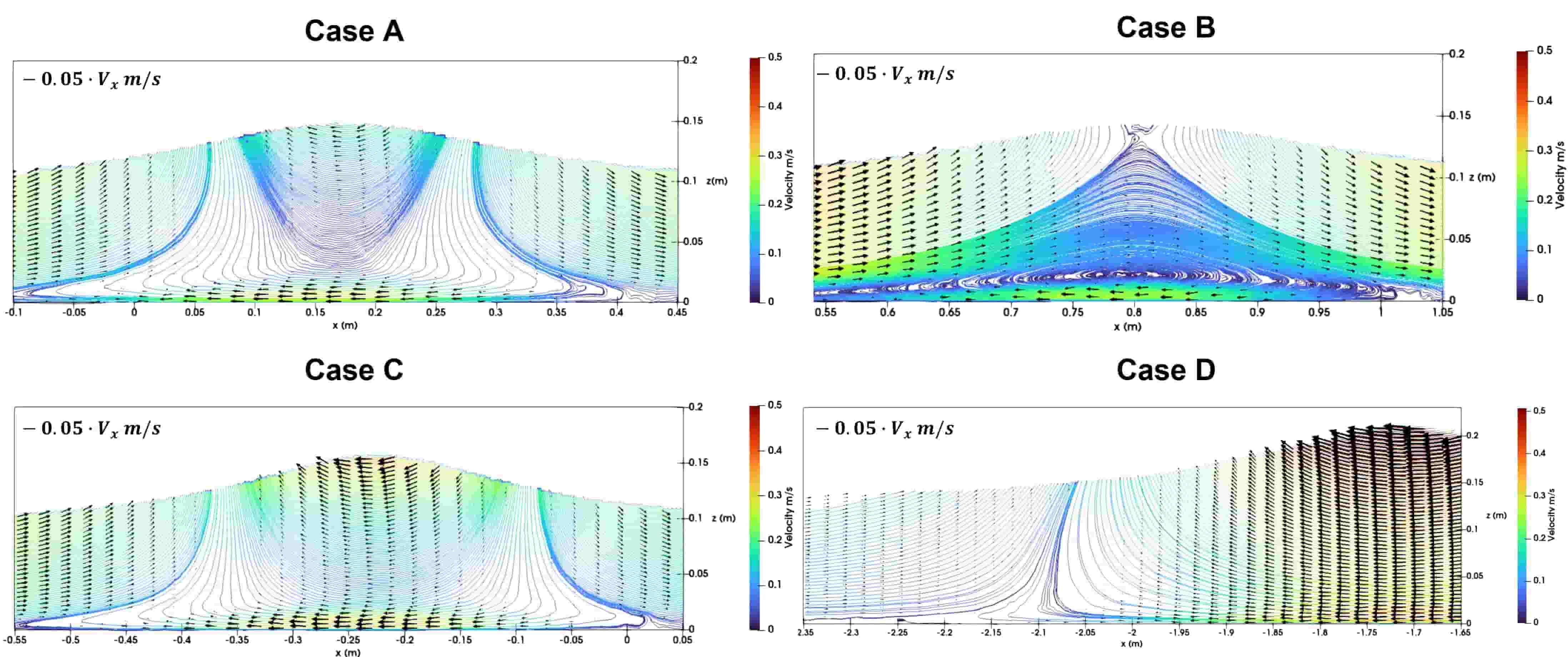}
	\caption{Spatial distribution of the velocity field beneath the bore front in the river bank reference frame $R_b$} 
for the four simulated cases A–D. Colors indicate the speed modulus with both downstream directed ($V_x>0$) and upstream directed ($V_x < 0$) flows. Case A exhibits a short‑lived local reversal confined to the vicinity of the front; Case B shows no reversal, with the river current remaining downstream oriented; Case C displays a mixed behaviour with a reversal at the front that persists downstream; and Case D reveals a full, persistent flow reversal throughout the water column. These patterns illustrate how the classical A–D classification depends on the observer’s reference frame.
\label{fig:berge}
\end{figure*}

Figure \ref{fig:berge} gives the spatial distribution of the flow velocity $V_{x_{|_{R_b}}}$ under the wave front in the reference frame of the river bank $R_b$ for the four cases, A, B, C and D, in the numerical simulation. We can observe that the river current is not reversed after the passage of the frontal wave in case B. The river current remains downstream oriented, no flow reversal (see Fig. \ref{fig:berge}  - top right). In case A, the river current shortly reverses after the passage of the wave front, but remains generally downstream. It is a local flow reversal (see Fig. \ref{fig:berge} - top left). In case C, the river current reverses at the front and remains reversed afterwards. It is a mixed between a local and global flow reversal (see Fig. \ref{fig:berge} - bottom left). In case D, the river current reverses permanently throughout the river. It is a global flow reversal (see Fig. \ref{fig:berge} - bottom right).

\begin{figure*}
\centering
\includegraphics[scale=0.1]{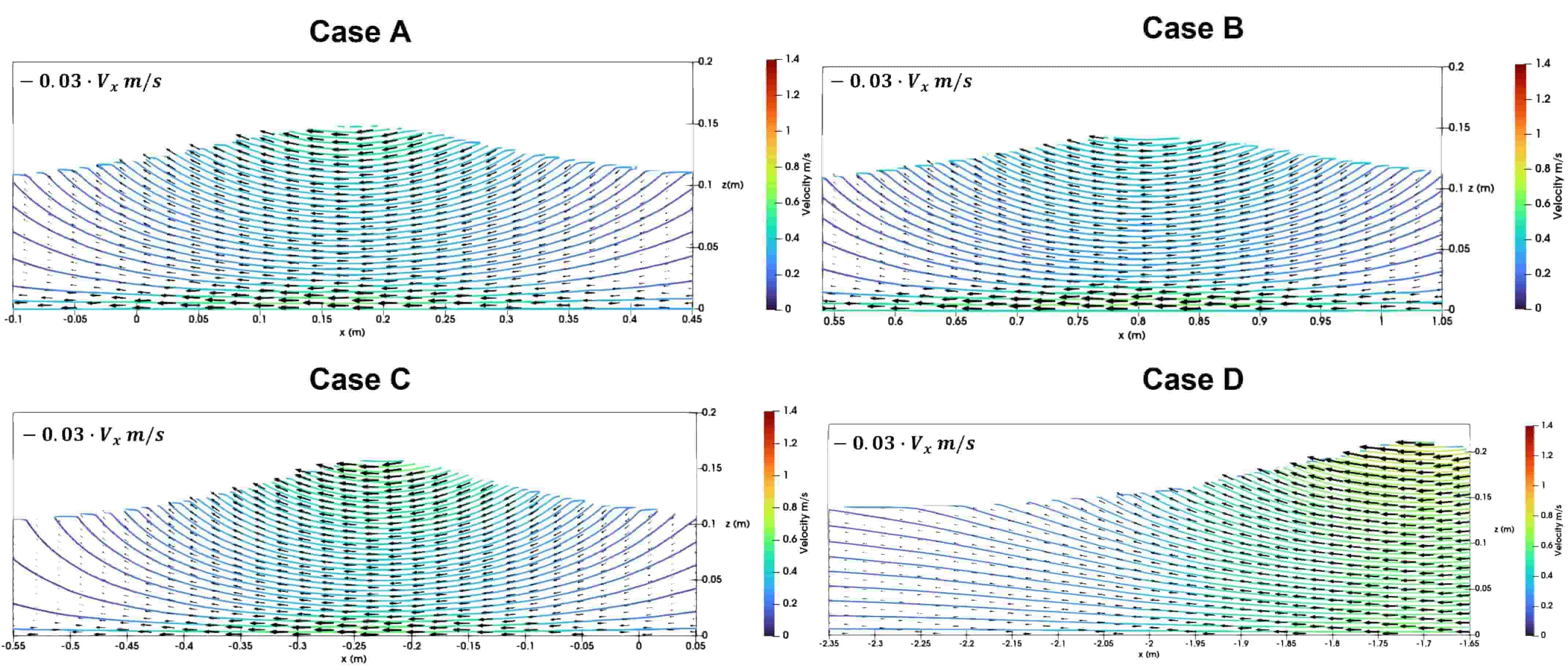}
	\caption{Spatial distribution of the velocity field beneath the bore front in the river current reference frame $R_c$ 
for the four simulated cases A–D. Transforming to $R_c$ reveals that a flow reversal occurs in all four cases, including Case B where no reversal is visible in the river bank frame. The reversal intensity and spatial extent vary markedly between cases, with Case D exhibiting a full, column‑wide reversal. In this frame, the background current is stationary, and the bore‑induced velocity dominates the dynamics.}
\label{fig:poisson}
\end{figure*}

Figures \ref{fig:poisson} and \ref{fig:surfeur} give the spatial distribution of flow velocities, $V_{x_{|_{R_c}}}$ and $ V_{x_{|_{R_m}}}$, under the wave front in the reference frames of the river current $R_c$ and the tidal bore $R_m$ for the four cases, A, B, C and D, in the numerical simulation. In the reference frame of the river current $R_c$, the flow reversal is now observable for the four cases. The flow reversal is global in case D and local for the other cases. In the reference frame of the river current $R_c$, the river current is perceive as still. The tidal bore seems to be moving up the river faster. In the reference frame of the  bore front $R_m$, the flow reversal is not observed in all the cases studied. In the reference frame of the bore front $R_m$, the tidal bore is perceive as still. The river current seems to be flowing faster.

\begin{figure*}
\centering
\includegraphics[scale=0.15]{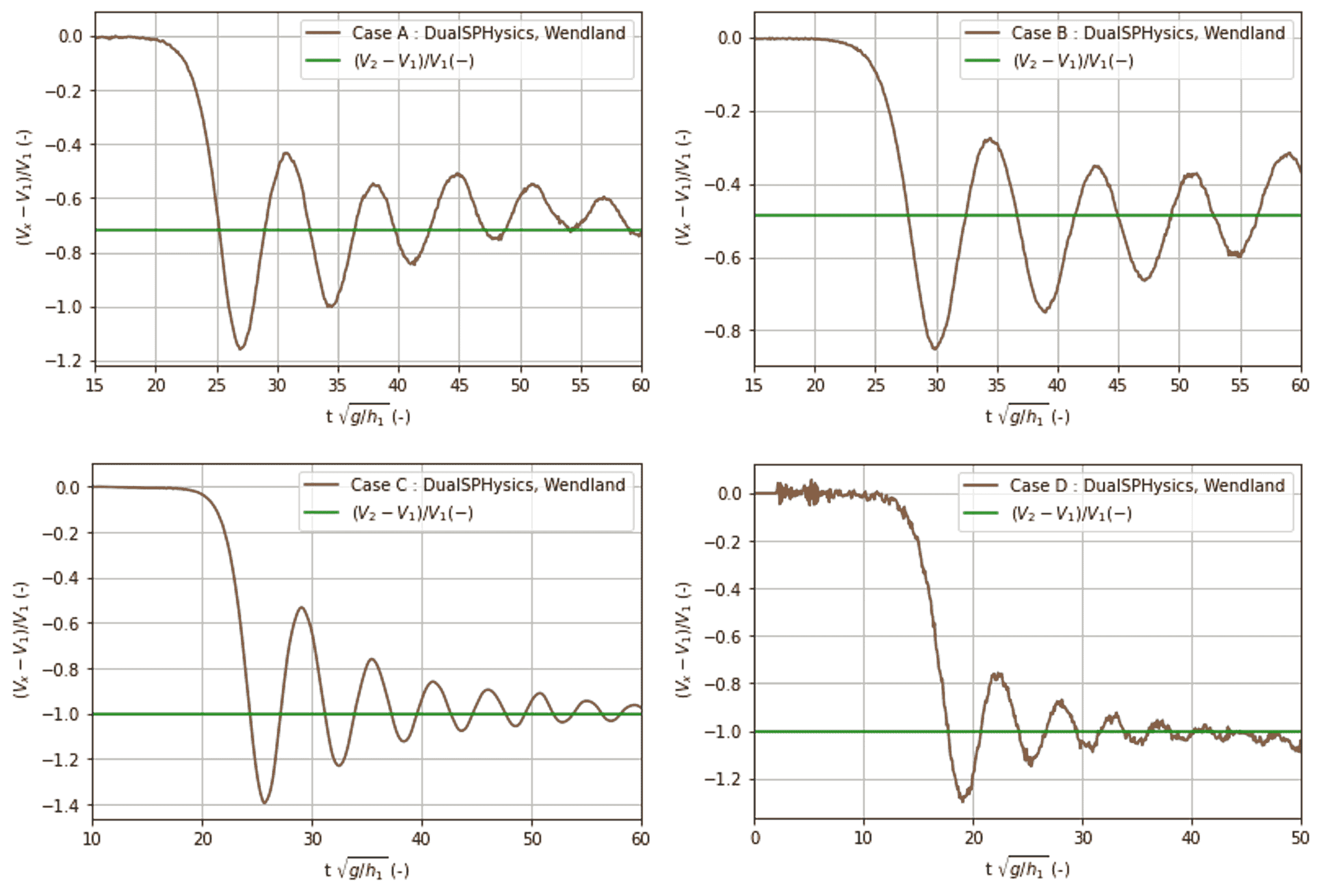}
	\caption{Non‑dimensional horizontal velocity $V_x/V_1$ in the river current reference frame $R_c$
for the four simulated cases A–D. For all cases, including Case B, which shows no reversal in the river bank frame, the transformation $V_{x_{|_{R_c}}} = V_{x_{|_{R_b}}} -V_1$ reveals a clear upstream directed region at the bore front. The reversal intensity increases from cases A to D, culminating in a fully developed, column‑wide reversal in Case D. This shows that, in the river current frame, the undular bore is intrinsically reversing regardless of its appearance in the bank frame.}
\label{fig:poisson_vit}
\end{figure*}

\begin{figure*}
\centering
\includegraphics[scale=0.1]{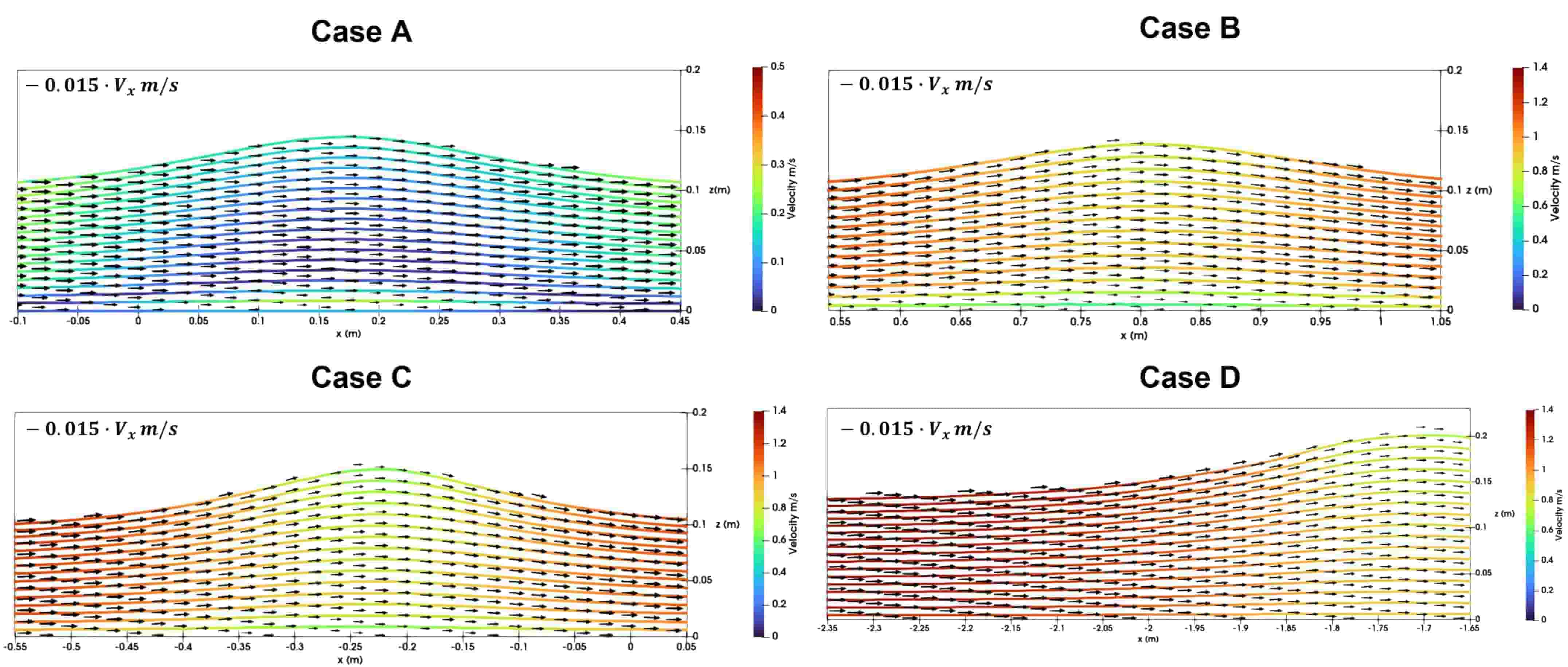} 
\caption{Spatial distribution of the velocity field beneath the bore front in the bore front reference frame $R_m$
for the four simulated cases A–D. In this frame, the bore front is stationary and the river flows toward it. No flow reversal is observed in any of the cases, in contrast with the river bank and river current frames. The relative structure of the bore becomes clearer: the front behaves as an hydraulic jump, and the downstream undulations appear as dispersive waves generated behind the jump.}
\label{fig:surfeur}
\end{figure*}

\begin{figure*}
\centering
\includegraphics[scale=0.7]{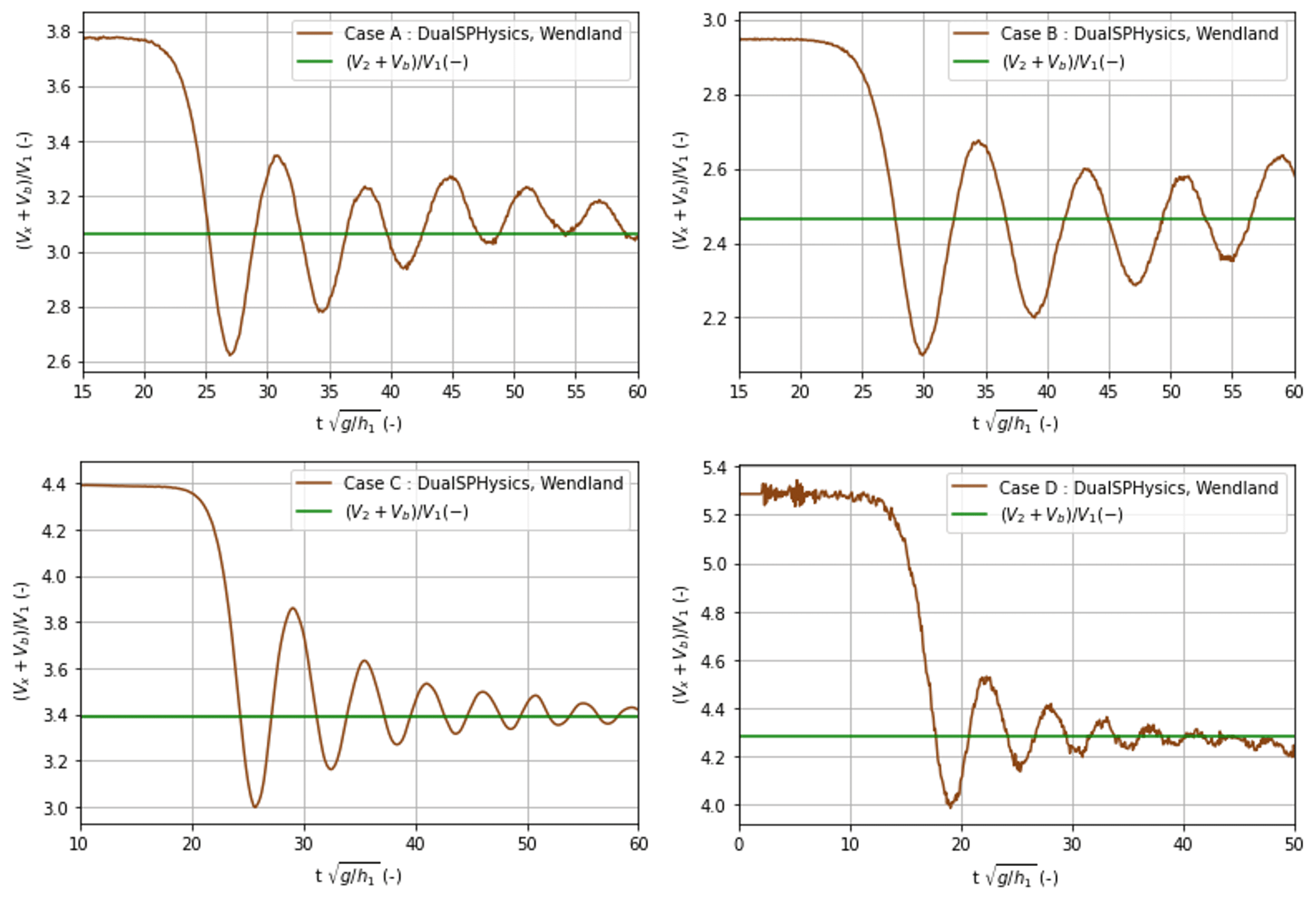}
	\caption{Non‑dimensional horizontal velocity $V_x/V_1$ in the bore front reference frame $R_m$
for the four simulated cases A–D. In this frame, where the bore front is stationary, no flow reversal is observed in any case. The velocity remains directed towards the bore, and the oscillations associated to the whelps occur entirely in the downstream directed region. This illustrates that, although flow reversal may appear in the river bank or river current frames, the intrinsic bore structure in $R_m$ is always non‑reversing as testified by the bore surfers.}
\label{fig:surfeur}
\end{figure*}


As indicated by Doyle and Ensign (2009) \cite{doyle2009}, the reference frame of the river bank $R_b$ is an Eulerian reference frame. This allows observation of the effects on the river bank, and is used for fixed measurements and human observations. The two other reference frames, considered in this work, are mixed Eulerian Lagrangian reference frames of an object's movement through its environment \cite{goodwin2006}. The reference frame of the bore front $R_m$ fixes the front of the tidal bore. This reference frame allows analysis of current disturbances caused by the tidal bore and is useful for modelling hydrodynamic interactions. The reference frame of the river current $R_c$ fixes the river current. This allows the study of the structure of the wave and local variations in the current to be studied, and is ideal for modelling local reversal.

\section{Conclusions}
In this work, the generation of tidal bores was studied numerically using the software DualSPHysics based on the WCSPH method with the partially closed gate method. The numerical study focused on undular tidal bores with a Froude number $Fr$ between 1.22 and  1.24. Experimentally, the tidal bore was generated on the rectangular open channel of the Institute Pprime with the partially closed gate method or the pumping stop method (1.25 $\leq Fr \leq 1.58$), the bores being weakly breaking that does not forbid us to look to the flow classification that would have been more complicated to report if the bores were fully breaking. The numerical results were first compared to the experimental data (see the subsection Model validation given in the section Appendices). The results showed good agreement on elevation of the water depth and the velocity profiles following the horizontal and vertical velocity components. The validated model was subsequently used to reproduce the four cases observed in the work by Berchet $et \, al .$ in 2018 \cite{berchet2018flow}. This previous study focused on the flow classification in the river bank frame of reference
. The four cases, denoted A, B, C and D, were reproduced experimentally using two methods of generation (partially closed gate and pump stop methods). In the numerical simulations, the cases were produced using the partially closed gate method only. 

The reference frame plays a role in the analysis of different phenomena related to the tidal bore. Depending on the reference frame, river bank, river current or bore front references, the flow reversals is observed or not:
\begin{itemize}
	\item The classification of Berchet $et \, al .$ was observed in the river current viewpoint. The four cases, A, B, C and D, were observed in both the numerical simulations and the experiments. Based on the velocity of the horizontal component, the reversal is local. As for Case B, no reversal is observed in this reference frame. Depending on the case, the streamlines shows the appearance of some structures as the elephant foot or a recirculation bubble.
		The Froude number $Fr$ is defined in the river current reference frame $R_c$. But in the river bank frame $R_b$, one can define other Froude numbers, $Fr_{bore|R_b} = V_b/\sqrt{gh_2}$ for the tidal bore and $Fr_{river current|R_b} = V_1 /\sqrt{gh1}$ for the river current. The ratio between the two Froude numbers is $V_b/V_1 \cdot \sqrt(h_1/h_2)$. The numerical simulations were simulated with a particularity. The fours cases, A, B, C and D, have the same values of $h_1$ and $h_2$. Thus, the ratio between the two Froude numbers, tidal bore and river current, is given by $V_b/V_1$. The reversal of undular and weakly breaking bores seems to be controlled by a specific kinematic condition. With regard to numerical results, the authors have identified the conditon following $V_b/V_1 > 2$. This identification is only done on 4 numerical simulations. A complete parametric study has to be performed.
\item In the river current reference frame $(V_{x_{|_{R_c}}} = V_{x_{|_{R_b}}} - V_{1} )$, the reversal of the tidal bore is observed in all the cases, notably in case B, where no previous reversal was observed except near the bed area. The reversal is thus observed in all cases but with different intensities of the first velocity component. In this frame of reference, the four cases become one, i.e, the tidal bore is always a reversing one.
\item The third reference frame; the reference frame of the bore front does not show a reversal in all the cases. The observer, travelling with the bore velocity, always sees the velocity following the river direction.  
\end{itemize}

This study provides a unified interpretation of tidal bore dynamics by explicitly revealing the central role of the observer’s reference frame. Combining time resolved PIV measurements in a laboratory flume with validated WCSPH numerical simulations, we reproduced the four flow regimes traditionally described in the theoretical and numerical literature, ranging from no reversal to full reversal, and demonstrated that they correspond to intrinsically different hydrodynamic states coupled to a kinematic projection. The classification in the numerical simulations is obtained using a constant water depth after the bore passage $h_2$ for simplification and so we can classify the different flow regimes. 
 This leaves the need to study the classification with an imposed varying in time range as in the field. This will allow to confirm the influence of the downstream depth in the channel or the post-bore depth in the estuary on the persistence of reversal.

In the river bank frame, historically used for nearly all observations, the apparent presence or absence of flow reversal depends on the competition between the river current $V_1$ and the bore front velocity $V_b$. Depending on their relative magnitude, the bore may appear non‑reversing, locally reversing, partially reversing, or fully reversing. By contrast, in the river current frame, every undular bore exhibits a reversal at the front, whereas in the bore front frame no reversal is observed. These results show that the classical classification of tidal bores primarily reflects the choice of reference frame in addition to genuinely distinct dynamical regimes A, B, C and D.

This Galilean relativistic interpretation provides a coherent basis for analysing tidal-bore behaviour in natural estuaries, where spatial variability in bathymetry, tidal range, and channel geometry may induce transitions between apparent regimes during a single propagation event as demonstrated here.

In summary, a universal description of tidal bores requires explicit specification of the measurement frame of reference. Integrating this aspect enables consistent interpretation of past studies and enhances the reliability of future analyses related to the interaction between tidal waves and river flows in high‑range estuarine systems.

\subsection*{Acknowledgments}
This work pertains (namely, is not funded but enters in the scientific perimeter) to the French government programs ``Investissement d'Avenir'' EUR INTREE (reference ANR-18-EURE-0010) and LABEX INTERACTIFS (reference ANR-11-LABX-0017-01). It has been funded by the ACI Pprime BABA in 2023-2024 (Pprime institute's internal funding scheme). It is currently funded by the french national research agency ANR throught the BABA program (ANR-25-CE01-3358). We thank the Plateforme Hydrodynamique Environnementale (PHE) from the Poitiers University which rented to us the open water channel Sedhyde available for the experiments. The authors would like to thank Patrick Braud, now retired metrological engineer, for his help in the course of the PIV settings and measurements. We would like to thank Alexis Bossard for his help in the course of the LED lighting experiments leading to the Figure 1.

\begin{appendices}

\label{Supplemental material}
\subsection{Appendix Model validation}
To describe accurately the cases of the reversal flow in undular tidal bores, the 2D (Two Dimensions) numerical model, developped with FreeCAD [\url{https://www.freecad.org/}] and DualSPHysics [\url{https://dual.sphysics.org/}], was tested on the generation of an undular tidal bore in an open channel using the partially or fully closed gate method. For the comparison, the experimental data were measured with the PIV (Particle Image Velocimetry) method \cite{thomas2022}.

The 2D numerical simulations were performed with a domain $L = 12\,m$ long (from $x=-5\,m$ to $x=7\,m$) and $h_1=0.094\,m$ high. The values of density and viscosity were fixed to $\rho_0 = 1000 \,kg/m^{3}$ and $\nu_0 = 10^{-6}\,m^{2}/s$ for water, respectively. The initial flow velocity was set uniform with $V_1 = 0.62\,m/s$ to reproduce an undular tidal bore with a Froude number $Fr=1.25$. The numerical simulations included a gate placed at position ($x=11\,m,z=0.15\,m$) going down with a velocity $V_g = 0.5\,m/s$ for $t_g=0.3\,s$ leaving an opening $h_g = 3.8\,cm$. The ratio $h/dp$, with $h$ the smoothing parameter and $dp$ the particles spacing, was fixed to $2$ for the Wendland (Eq. (\ref{wendland})) and Cubic B-spline (Eq. \ref{cubic}) functions, classically used in the scientific literature \cite{jandaghian2022} \cite{monaghan1985} \cite{wendland1995}. For the predictor-corrector scheme used, a time step $\delta t_{min} = 8.10^{-5}\,s$ was set \cite{dominguez2022} \cite{heydemans2024}. The 2D numerical simulations were performed for a real time $t=7\,s$. It was done using GPU (Graphical Processing Unit) computing on a computer equipped with the graphics card NVIDIA Corporation GK 104 [GeForce GTX 660 Ti] (rev a1), the language CUDA (Compute Unified Device Architecture) version 11.4 and the system management interface NVIDIA 470.256.02. The GPU time was approximately $20$ minutes for each numerical simulation \cite{dominguez2022} \cite{heydemans2024}. The experimental data represent one run instead of an averaged data which is generally the case in the scientific literature \cite{berchet2018flow}.

\begin{figure}
\centering
\includegraphics[width=0.4\textwidth]{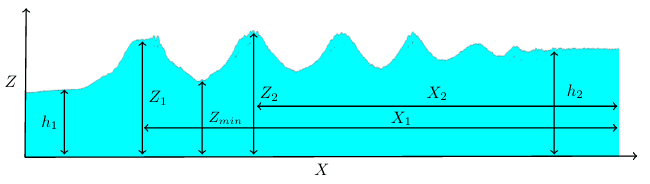};
\caption{Definition sketch of the geometric quantities used to evaluate the wave amplitude $a_w$ and wavelength $L_w$ of an undular bore, as given in Eq. \ref{eq:comparison}. The front elevation $Z_1$, minimum elevation $Z_{\min}$ between the front and the first whelp, and downstream elevation $Z_2$ are indicated, as well as the corresponding horizontal positions $X_1$ and $X_2$. The still water depths upstream and downstream of the bore, $h_1$ and $h_2$, are also shown.}
\label{fig:amplitude}
\end{figure}

The convergence of the WCSPH (Weakly Compressible Smoothed Particle Hydrodynamics) method with respect to the particles spacing $dp$ was studied. The particles spacing $dp$ was varied between 0.002 $m$ and 0.006 $m$. The smoothing length $h$ was then estimated by guaranteeing a ratio ${h}/{dp}=2$. To estimate the accuracy of the WCSPH method, we have compared the numerical results to experimental data by observing two physical quantities, the wave amplitude $a_w$ and the wavelength $L_w$ of the undular tidal bore. These two physical quantities were extracted from the numerical results and the experimental data by using paraview to identify the free surface (see Fig. \ref{fig:amplitude}). Their values are given by:
\begin{equation}
a_w = \frac{Z_1 - Z_{min}}{2}
\quad \mbox{and} \quad
L_w = X_1 - X_2 
\label{eq:comparison}
\end{equation}
where $Z_1$ is the front elevation, $Z_{min}$ is the lowest height between the front and the first whelp, $X_1$ and $X_2$ are the distance from the starting point of propagation of the bore (right) to the front, first whelp, respectively (see Fig. \ref{fig:amplitude}).

\begin{figure}
\centering
\includegraphics[width=0.45\textwidth]{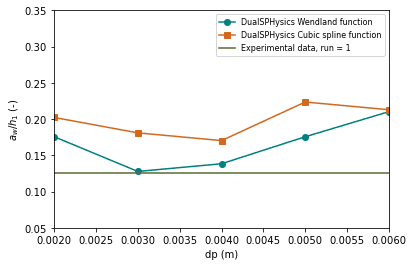}
\includegraphics[width=0.45\textwidth]{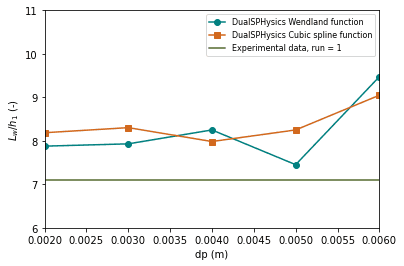}
\caption{Dimensionless wave amplitude $a_w/h_1$ (top) and dimensionless wavelength $L_w/h_1$ (bottom) as functions of the particle spacing $dp$ for both SPH (Smoothed Particle Hydrodynamics) kernel functions used in the convergence study. Red squares denote results obtained with the Cubic B-spline kernel, and teal green circles denote results obtained with the Wendland kernel. The olive green horizontal line indicates the experimental reference values $a_w/h_1 = 0.13$ and $L_w/h_1 = 7.09$. Both kernels converge for $dp < 0.004\,m$, with the Wendland kernel yielding closer agreement with the experimental measurements.}
\label{fig:aw}
\end{figure}

Figure \ref{fig:aw} presents the non dimensional values of wave amplitude $a_w/h_1$ (Left) and wavelength $L_w/h_1$ (Right), as functions of the distance between the particles $dp$, for both the Cubic B-spline and Wendland functions. The five values obtained with the Cubic B-spline and Wendland functions are presented by red squares and blue circles, respectively. The green line presents the experimental value of two physical quantities studied, $a_w/h_1 = 0.13$ and $L_w/h_1 = 7.09$. From Fig. \ref{fig:aw}, we can observe that we have a convergence of two kernel functions to $dp < 0.004\,m$ for both physical quantities studied. The Wendland function converges to the values of $a_w/h_1 = 0.18$ and $L_w/h_1 = 7.9$. The Cubic B-spline function converges to the values of $a_w/h_1 = 0.21$ and $L_w/h_1 = 8.2$. We also observe that the Wendland kernel function is closer to the experimental value.

A comparison of velocity profiles between the numerical results and the experimental data was done on both of two flow velocity components, horizontal $V_x$ and vertical $V_z$. The velocity profiles were obtained from the time evolution of each flow velocity component at a given position in the open channel. The abscissa of the observation position was set to $X = 9\,m$ for the numerical results and $X = 0\,m$ for the experimental data (the middle of the observation window of experimental data). For both the numerical results and the experimental data, two values of the vertical elevation $Z$ of the observation position were set to 60 $mm$ (in the water column) and 10 $mm$ (near the bed). For the numerical results, we have taken a distance between the particles $dp=0.002\,m$ and a smoothing length $h=0.004$.

\begin{figure}
\centering
\includegraphics[width=0.45\textwidth]{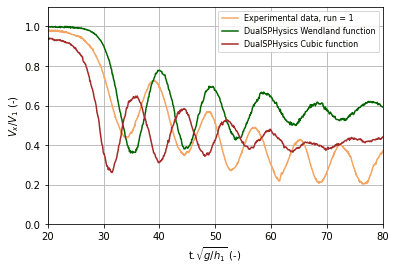}
\includegraphics[width=0.45\textwidth]{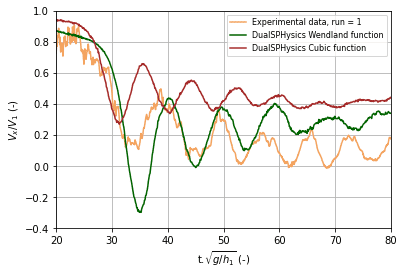}
	\caption{Non-dimensional value of the horizontal velocity component ${V_x}/{V_1}$ as a function of non-dimensional time $t \sqrt{g/h_1}$ for two values of the vertical elevation, $Z$ = 60$mm$ (Left) and 10 $mm$ (right). Non-dimensional horizontal velocity ${u_x}/{V_1}$ as a function of the non-dimensional time $t \sqrt{g/h_1}$ for two vertical elevations: $z = 60$ $mm$ (left) and $z = 10$ $mm$ (right). Experimental data are shown in orange, while WCSPH numerical results obtained with the Wendland kernel (green) and the Cubic B-spline kernel (red) are superposed. For both elevations, the Wendland kernel reproduces the experimental temporal evolution more accurately, particularly for $0 < t^* < 55$. The Cubic B-spline kernel predicts a faster bore passing, causing increasing phase discrepancy and signal loss at later times (beyond $t^* \approx 65$). The drop in the experimental signal near the end of the time series is attributed to the progressive lowering of the water level.}
\label{fig:60}
\end{figure}

Figure \ref{fig:60} presents the non-dimensional value of the horizontal velocity component ${V_x}/{V_1}$ as a function of the non-dimensional time $t^\ast = t \sqrt{{g}/{h_1}}$ for the two values of the vertical elevation, $Z = 60\,mm$ (top) and $10\,mm$ (bottom). The numerical results are represented in green for the Wendland function and in red for the Cubic B-spline function. The experimental data are represented in yellow. For the two values of $Z$, we can observe that the Wendland function succeeds to have the same shapes as the experimental data for $0 < {t^\ast} < 55$, unlike the Cubic B-spline function. The Cubic B-spline has a velocity of propagation $V_b$ bigger and the signal is just lost at the end (starting from the value of almost $t^*$ = 65). For $Z$ = 60 $mm$, we can see that the signal of experimental data started dropping from ${t^\ast} = 55$ as the water level in the channel was too small. For the two values of $Z$, the oscillations of the experimental data are significant since the data represents one run.

\begin{figure}
\centering
\includegraphics[width=0.45\textwidth]{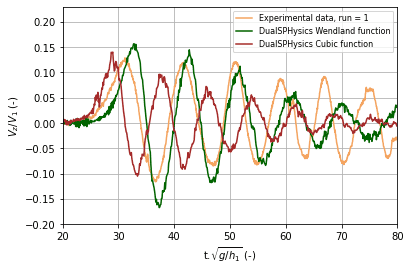}
\includegraphics[width=0.45\textwidth]{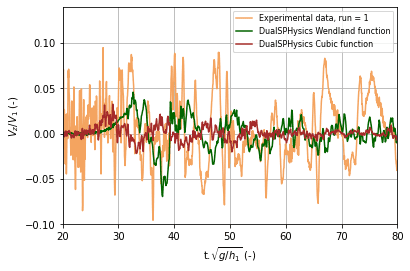}
	\caption{Non-dimensional value of the vertical velocity component ${V_z}/{V_1}$ as a function of non-dimensional time $t \sqrt{g/h_1}$ for two values of the vertical elevation, $Z$ = 60 $mm$ (left) and 10 $mm$ (right). Non-dimensional vertical velocity $V_z/V_1$ as a function of the non-dimensional time $t \sqrt{g/h_1}$ for two vertical elevations: $z=60\,mm$ (left) and $z=10\,mm$ (right). Experimental data are shown in orange, while WCSPH numerical results using the Wendland kernel (green) and the Cubic B-spline kernel (red) are superposed. At $Z=60\,mm$, the Wendland kernel reproduces the experimental temporal signal more faithfully, whereas the Cubic B-spline kernel again predicts a faster bore passing. At $Z=10\,mm$, strong oscillations in the experimental data arise from the reduced water depth near the bed. Overall, the Wendland kernel shows better agreement with the measured vertical velocity dynamics.}
\label{fig:10}
\end{figure}

	Figure \ref{fig:10} presents the non-dimensional value of the vertical velocity component ${V_z}/{V_1}$ as a function of the non-dimensional time $t^\ast$ for the two values of the vertical elevation, $Z = 60\,mm$ (top) and $10\,mm$ (bottom). The numerical results are represented in green for the Wendland function and in red for the Cubic B-spline function. The experimental data are represented in yellow. For $Z = 60 \,mm$, we can observe the same behavior than for the horizontal component $V_x$ of the velocity flow. The Cubic B-spline function gives a faster bore passing. The Wendland function succeeds to have the same shape as the experimental data. For $Z = 10\, mm$, we can see that the oscillations of the experimental data were so strong due to the same fact cited before.

\begin{figure}
\centering
\includegraphics[width=0.7\columnwidth]{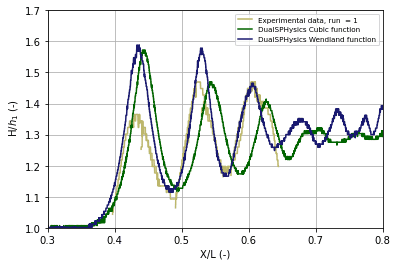}
\caption{Non-dimensional free surface elevation $H/h_1$ as a function of the non-dimensional horizontal position $x/L$, comparing WCSPH numerical simulations with experimental measurements. Experimental data are shown in olive green, while numerical results using the Wendland kernel (blue) and the Cubic B-spline kernel (green) are superposed. Both kernels overestimate the wave front elevation relative to the experiment; however, the Wendland kernel reproduces more accurately both the instant and the amplitude of the first whelp, yielding closer agreement downstream. The Cubic B-spline kernel fails to capture the subsequent undulations beyond $x/L \approx 0.65$.}
\label{fig:sl125}
\end{figure}

A comparison of the free surface between the numerical results and the experimental data was also done. For both the numerical results and the experimental data, the free surface was extracted with Paraview [\url{https://www.paraview.org/}]. The experimental data are represented in yellow. It is required to note that the PIV technique is originally used for the values of the flow velocity and it is not used to have the free surface. Here, we wanted to have an idea about the difference between the numerical results and the experimental data. Figure \ref{fig:sl125} presents the non-dimensional value of the water depth ${H}/{h_1}$ as a function of the non-dimensional value of the position ${x}/{L}$ for both the numerical results and the experimental data. The numerical results are represented in green for the Cubic B-spline function and in blue for the Wendland function. The experimental data are represented in yellow. We can observe that the wavefront elevation obtained with the two kernel functions is higher compared to the experimental one. Unlike the Cubic B-spline function, the Wendland function has the same timing as the experimental data which will kept for the first whelp, this time with almost the same elevation to have a perfect match after. The Cubic B-spline function doesn't succeed to catch the other undulations starting from the value of $x/L = 0.65$.

\newpage
\subsection{Appendix Metrological Setup}

The water channel is equipped with a LED lightning system (White LED on 3m long (24V, 6000K)) for flow visualizations and ultrasonic sensors mic+25/D/TC from Microsonic to measure the water depths along the water channel. The range is between $30-250mm$ and the resolution of the sensor is $25\mu m$.

\begin{figure}
\centering
\includegraphics[width=0.4\textwidth]{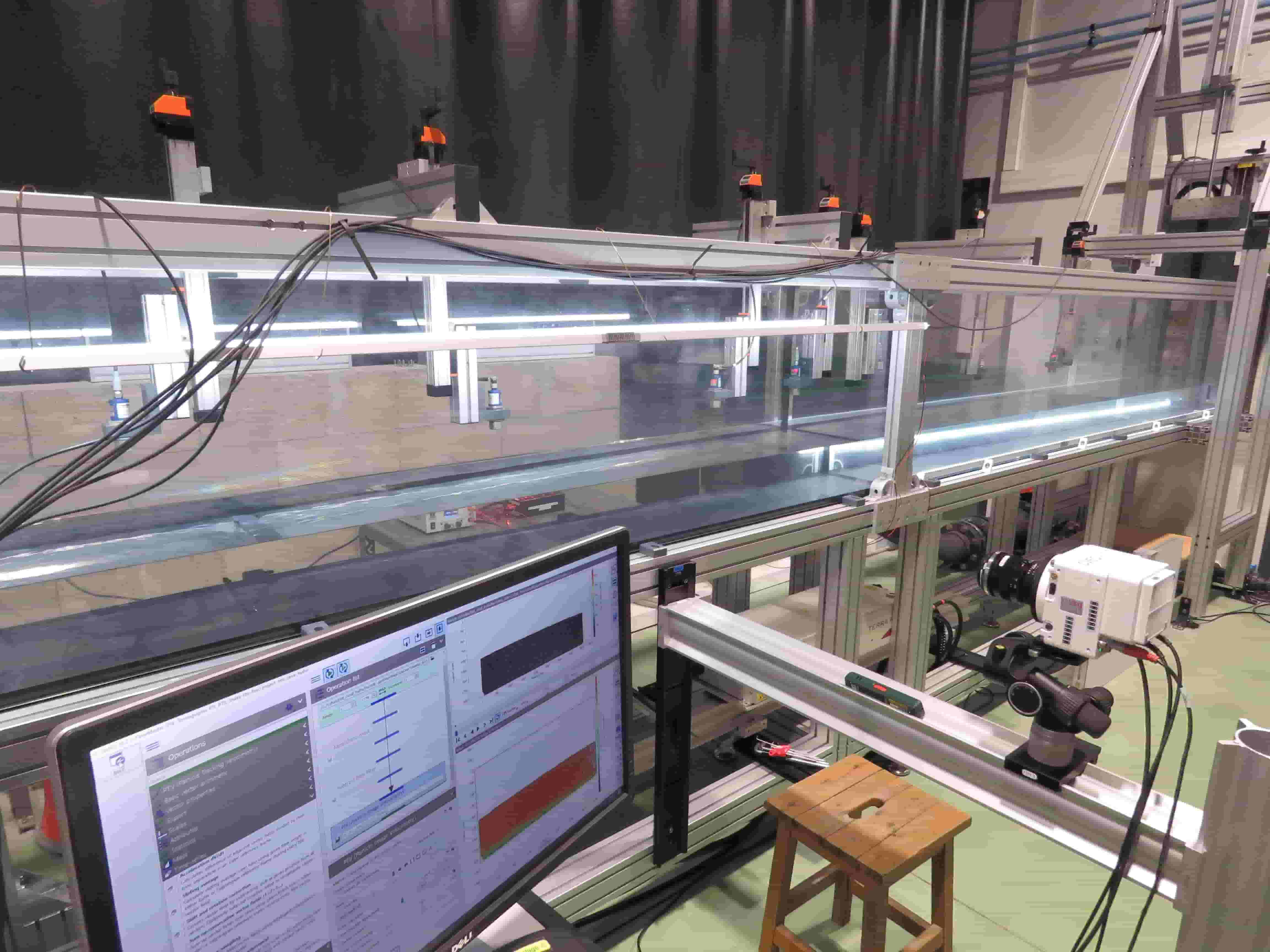}
\caption{LED lighting system and lateral PIV camera used in the experimental setup. Fluorescent seeding particles are illuminated by the laser sheet, while several Microsonic ultrasonic sensors positioned above the channel monitor the mean water depth. The flow direction is from left to right.}
\label{EXP}
\end{figure}

PIV (Particle Image Velocimetry) measurements were performed with fluorescent particles as PIV tracers made of polyamid with relative density $1.04$ and $20 \mu m$ of diameter doped with a fluorescent dye rhodamines 6G whose pink signal is separated from the green LASER light by a high-pass filter (see Fig. \ref{EXP}). Hence, spurious reflections are dismissed. As usual in PIV measurements, we lack some signal close to the free surface that is masked in a subsequent post-processing.

Tidal bore regimes are probed experimentally using 2D two components PIV measurements time-resolved method (2C-2D TR-PIV) that records two components of the velocity on a whole vertical plane. This allows the estimation of flow topology and distribution of the bi-dimensional averaged velocity throughout the entire water column.

\begin{figure}
\centering
\includegraphics[width=0.4\textwidth]{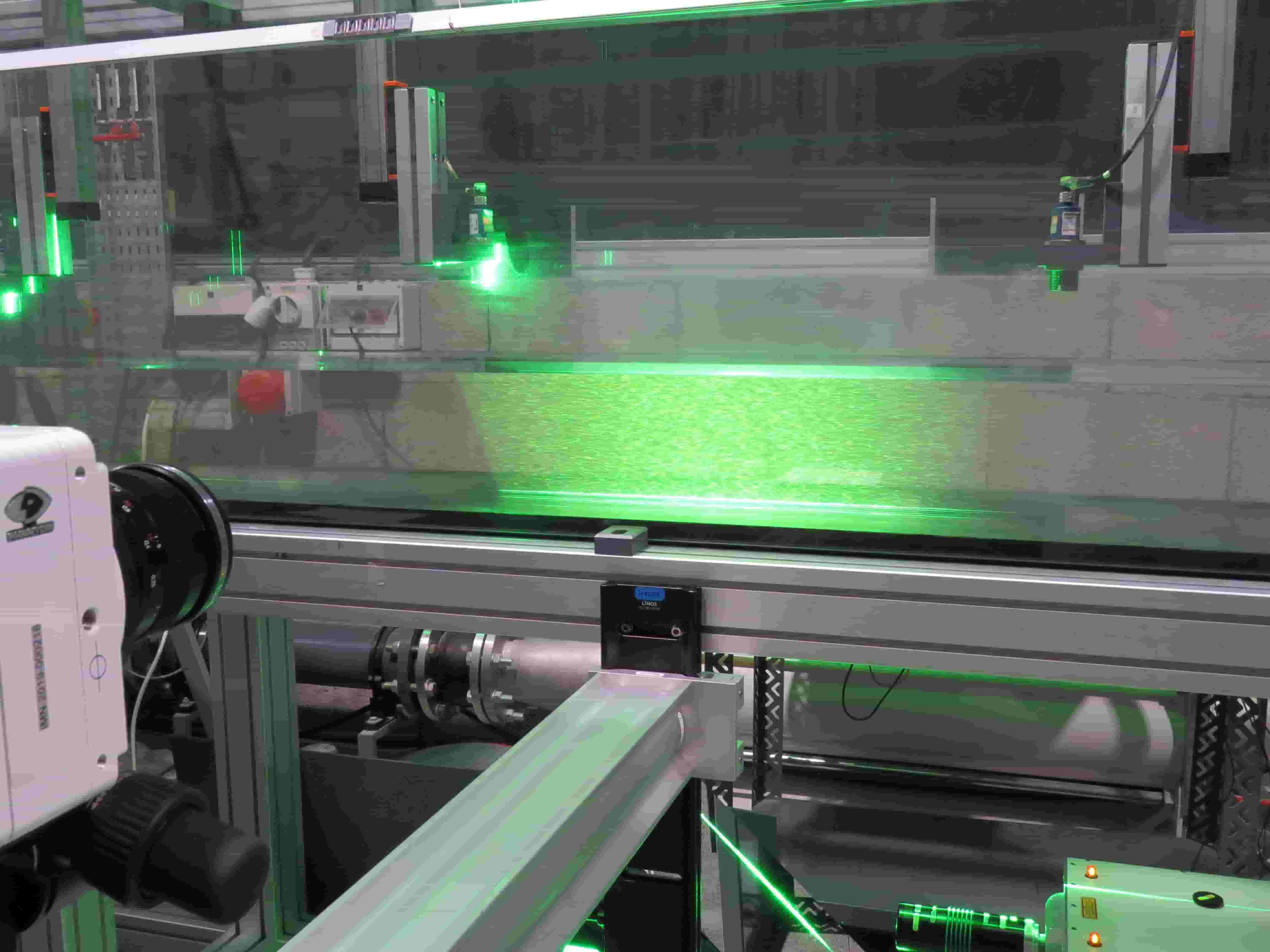}
\caption{Lateral view of the LASER sheet projected upward through the transparent channel bottom using a mirror and dedicated optics. Fluorescent PIV particles appear as green dots where they intersect the laser plane. The background flow proceeds from left to right.}
\label{PIVsetup}
\end{figure}

We used a manual trigger then recorded the particles displacement in the water column typically 12 s of duration (3000 images) at 250Hz ($4\,ms$ between two images, noted as $im$) with a Phantom VEO 4K 990L Camera (8 MegaPixels at $1 kHz$) equipped of a 50 mm Nikon lens and a high-pass filter at $540\, nm$ in optical wavelength. The flow was illuminated by a  $2 \times 30\, mJ$ pulses Nd-YLF Continuum Terra PIV (optical wavelength $527 \,nm$) within a plane parallel to the open water channel located at the center by a mirror from below (see Fig. \ref{PIVsetup}). Velocity vectors and related streamlines colorized by the speed modulus were computed using the software Davis 8 from Lavision. In particular, the PIV post-processing involved three passes: the first two passes are performed with a $64\times 64$ pixels interrogation window, which is decreased to $32\times 32$ pixels with 50\% overlap for the final pass. 2500-4000 instantaneous fields at 250 Hz were recorded for each experimental case, which is sufficient for statistical estimations.

\end{appendices}

\bibliography{ILCS_biblio}

\end{document}